\documentclass[12pt]{elsarticle}

\makeatletter
\def\ps@pprintTitle{%
 \let\@oddhead\@empty
 \let\@evenhead\@empty
 \def\@oddfoot{\centerline{\thepage}}%
 \let\@evenfoot\@oddfoot}
\makeatother

\usepackage{hyperref}
\usepackage{amsmath,amsfonts,amssymb}
\usepackage{url}
\usepackage{bm}
\usepackage{listings}
\usepackage{color}

\usepackage[absolute]{textpos}
\usepackage{braket}

\usepackage{etoolbox}
\patchcmd{\pprintMaketitle}
  {\hrule\vskip12pt}
  {\hrule\vskip12pt\ifvoid\extrainfobox\else\unvbox\extrainfobox\par\vskip12pt\fi}
  {}{}

\newsavebox\extrainfobox

\newcounter{bla}

\newcommand{\bfp}{\ensuremath{\mathbf{p}}}
\newcommand{\bfk}{\ensuremath{\mathbf{k}}}
\newcommand{\bfA}{\ensuremath{\mathbf{A}}}
\newcommand{\bfV}{\ensuremath{\mathbf{V}}}
\newcommand{\bfx}{\ensuremath{\mathbf{x}}}
\newcommand{\bfy}{\ensuremath{\mathbf{y}}}
\newcommand{\bfz}{\ensuremath{\mathbf{z}}}

\newcommand{\bfq}{\ensuremath{\mathbf{q}}}
\newcommand{\bfD}{\ensuremath{\mathbf{D}}}
\newcommand{\bfSigma}{\ensuremath{\bm{\sigma}}}
\newcommand{\bfEps}{\ensuremath{\bm{\epsilon}}}
\newcommand{\bfDel}{\ensuremath{\mathbf{\nabla}}}

\begin{document}

\begin{frontmatter}

\title{New Developments in FeynCalc 9.0}

\author[a]{Vladyslav Shtabovenko\corref{author1}}
\author[b]{Rolf Mertig\corref{author2}}
\author[c]{Frederik Orellana}

\cortext[author1] {\textit{E-mail address:} v.shtabovenko@tum.de}
\cortext[author2] {\textit{E-mail address:} rolfm@gluonvision.com}

\address[a]{Technische Universit\"at M\"unchen, Physik-Department T30f, James-Franck-Str. 1, 85747 Garching, Germany}
\address[b]{GluonVision GmbH, B\"otzowstr. 10, 10407 Berlin, Germany}
\address[c]{Technical University of Denmark, Anker Engelundsvej 1, Building 101A,
2800 Kgs. Lyngby, Denmark}

\begin{textblock*}{30ex}(\textwidth,5ex)
TUM-EFT 71/15
\end{textblock*}

\begin{abstract}
In this note we report on the new version of \textsc{FeynCalc}, a \textsc{Mathematica} package for symbolic semi-automatic evaluation of Feynman diagrams and algebraic expressions in quantum field theory. The main features of version 9.0 are: improved tensor reduction and partial fractioning of loop integrals, new functions for using \textsc{FeynCalc} together with tools for reduction of scalar loop integrals using integration-by-parts (IBP) identities, better interface to \textsc{FeynArts} and support for $SU(N)$ generators with explicit fundamental indices.

\end{abstract}
\begin{keyword}
High energy physics; Feynman diagrams; Loop integrals; Dimensional regularization; Dirac algebra; Color algebra; Tensor reduction;

\end{keyword}

\end{frontmatter}

\section{Introduction}
In the last decades the importance of computer tools for higher order perturbative calculations in quantum field theory (QFT) has increased tremendously.
Indeed, some recent achievements \cite{Anastasiou2015,Marquard2015, Beneke2014a} in this field would have been hardly possible to complete within a reasonable time frame, if such projects were to be carried out only by pen and paper. The question that most QFT practitioners pose themselves today is not whether to use software tools or not, but rather which combination of tools will be the most efficient for the endeavored project. It is clear that, in principle, there can be no universal package to cover any demand of any particle theorist. Instead, specific programs that provide different level of automation should be used for specific tasks. One of such specific tools is \textsc{FeynCalc} \cite{Mertig1991}, that recently was released in the version 9.0.

\textsc{FeynCalc} is a \textsc{Mathematica} package for algebraic calculations in QFT and semi-automatic evaluation of Feynman diagrams. The very first public version of \textsc{FeynCalc}, developed by Rolf Mertig with guidance from Ansgar Denner and Manfred B\"ohm, appeared in 1991. The main developments and improvements between 1991 and 1996 were triggered by the work of Mertig in electroweak theory \cite{Denner1992, Beenakker1993, Beenakker1994, Beenakker1993a} and perturbative QCD \cite{Mertig1996}.  In 1998 Rolf Mertig and Rainer Scharf released \textsc{TARCER} \cite{Mertig1998}, a \textsc{Mathematica} package that implements reduction of 2-loop propagator type integrals with arbitrary masses using O. V. Tarasov's recurrence algorithm \cite{Tarasov1996,Tarasov1997}. Since then \textsc{TARCER} is a part of \textsc{FeynCalc} that can be loaded on-demand. Between 1997 and 2000, important contributions to the project came from {Frederik} {Orellana}, who, besides working on the general code, contributed the sub-package \textsc{PHI} for using \textsc{FeynCalc} in Chiral Perturbation Theory (\(\chi\)PT) \cite{Buechler2001} and interfacing to \textsc{FeynArts} 3 \cite{Hahn2001}. From 2001 until 2014, with both developers out of theoretical physics, the development of \textsc{FeynCalc} was mostly constrained to bug fixing and providing support through the mailing list \footnote{\url{http://www.feyncalc.org/forum}}, although some interesting projects with external collaborators still were conducted \cite{Feng2012}. In 2014, the developer team was joined by Vladyslav Shtabovenko, who started to work on rewriting some parts of the existing code and implementing  new features. In the same year the source code repository of \textsc{FeynCalc} was moved to \textsc{GitHub} \footnote{\url{https://github.com/FeynCalc}}, where the master branch of the repository represents the current development snapshot of the package. Not only the stable releases, but also the development version of \textsc{FeynCalc} can be anonymously downloaded by everyone at any time free of charge. The code is licensed under the General Public License (GPL) version 3.
To minimize the number of new bugs and regressions, an extensive unit testing framework \footnote{\url{https://github.com/FeynCalc/feyncalc/tree/master/Tests}} with over 3000 tests was introduced.

This note is organized in the following way. Section 2 compares \textsc{FeynCalc} to other packages for automatic evaluation of 1-loop Feynman diagrams and discusses setups, in which \textsc{FeynCalc} can be particularly useful. Section 3 provides an overview of interesting new features and improvements in \textsc{FeynCalc} 9.0. Section 4 gives an example of using \textsc{FeynCalc} to determine matching coefficients in NRQCD \cite{Bodwin1995}, a non-relativistic effective field theory (EFT) for heavy quarkonia. Finally, we summarize and draw our conclusions in Section 5.

\section{Comparison to similar tools} \label{sec:comparison}

In view of the existence of several well-known symbolic packages (\textsc{FormCalc} \cite{Hahn1999}, \textsc{GoSam} \cite{Cullen2014}, \textsc{FDC} \cite{Wang2004},
 \textsc{GRACE} \cite{Belanger2006}, \textsc{Diana} \cite{Tentyukov1999}) that offer almost fully automatic evaluation of Feynman diagrams at 1-loop from Lagrangian till the cross-section, it appears necessary to explain how \textsc{FeynCalc} differs from such tools and why it is  useful.

\textsc{FeynCalc} by itself does not provide a fully automatic way of computing cross sections or decay rates.
Indeed, \textsc{FeynCalc} cannot generate Feynman diagrams and has no built-in capabilities for the numerical evaluation of master integrals and for the phase space integration. Therefore, these two important steps should be done using other tools.

Second, \textsc{FeynCalc} normally performs all the algebraic manipulations using \textsc{Mathematica}. This leads to a slower performance when compared to tools that rely e.g. on \textsc{FORM} \cite{Vermaseren2007} for the symbolics. Despite some possibilities \cite{Feng2012} to link \textsc{FeynCalc} with \textsc{FORM}, one should keep in mind that \textsc{FeynCalc} is not very well suited for evaluating hundreds, thousands or even millions of Feynman diagrams.

Finally, \textsc{FeynCalc} does not impose any particular ordering in which different parts (Dirac matrices, $SU(N)$ matrices, loop integrals etc.) of the amplitudes are supposed to be computed. It is always up to the user to decide what is the most useful way to carry out the calculation. This particular feature makes \textsc{FeynCalc} very different from tools that attempt to automatize all the steps of the evaluation process. Such tools usually stick to a particular workflow which roughly consists of the following steps:

\begin{enumerate}
\item The user specifies the process that needs to be computed.
\item  If the given process is available in the standard configuration, load the corresponding model (e.g. Standard model (SM)). Otherwise the user must create a new model that contains this process.
\item Using the loaded model, generate relevant Feynman diagrams for the given process.
\item Evaluate the amplitudes by performing all the necessary algebraic simplifications.
\item Square the amplitude and sum/average  over the spins of the involved particles.
\item Integrate over the phase space.
\end{enumerate}

In this list, already the second step might turn out to be problematic. The list of built-in models usually includes SM and some popular (e.g. SUSY inspired) extensions, while more exotic theories require custom model files to be added by the user. If the Lagrangian of such a theory looks very different from $\mathcal{L}_\textrm{SM}$ (e.g. 
in EFTs that are not strictly renormalizable (with an arbitrary number of legs in vertices) like \(\chi\)PT or even not manifestly Lorentz covariant like non-relativistic QCD (NRQCD) \cite{Bodwin1995} or potential non-relativistic QCD (pNRQCD) \cite{Brambilla2000}), then its implementation becomes a formidable task. On the other hand, even if the model can be implemented with a limited amount of effort, is still might cost more time than just writing down the amplitudes by hand and then manually entering them into the program. Although it is possible to make fully automatic tools accept such amplitudes as input, this is usually much less straight-forward than the standard way of just specifying the process, launching the diagram generator and letting the automatics do the rest.

\textsc{FeynCalc} avoids such difficulties by accepting any kind of input that consists of valid \textsc{FeynCalc} objects. Hence, one can enter e.g. standalone Dirac traces, Lorentz vectors or loop integrals and then manipulate them with suitable \textsc{FeynCalc} functions. In this sense \textsc{FeynCalc} can be used much like a ``calculator'' for QFT expressions.

For manual input of Feynman diagrams \textsc{FeynCalc} contains some functions (\texttt{FeynRule}, \texttt{FunctionalD}, \texttt{CovariantD}, \texttt{QuantumField} etc.) for deriving Feynman rules from Lagrangians that are manifestly Lorentz covariant. Furthermore, it is also possible to evaluate Feynman diagrams that were generated automatically (e.g. by \textsc{FeynArts}), so that the user always can choose the most efficient strategy to get the calculation done. 

Steps 4 and 5 usually imply that the user is not supposed to interfere too much with the evaluation process. Instead, one should rely on the available options to influence the outcome of the calculation. For example, when an automatic tool handles the Dirac algebra, it would normally try to simplify everything it can. While in general, this approach is perfectly fine, sometimes one would like to simplify only some of the Dirac structures, leaving the others (e.g. all the traces involving an odd number of $\gamma^5$) untouched. In principle, provided that the particular tool is open source, one can always modify its code accordingly to obtain the desired output. Depending on the complexity of the code and the amount of documentation, this might, however, take some time and even introduce new bugs.

With \textsc{FeynCalc}, the same result can be achieved in a more simple way, as one always has full access to all kind of intermediate expressions. For this purpose \textsc{FeynCalc} also provides various helper functions (e.g. \texttt{Collect2}, \texttt{Expand2}, \texttt{Factor2}, \texttt{Isolate}, \texttt{ExpandScalarProduct}, \texttt{DiracGammaExpand}, \texttt{MomentumCombine}, \texttt{FCLoopSplit}, \texttt{FCLoopIsolate}, \texttt{FCLoopExtract}) that can be used to expand, sort, abbreviate and collect the given expressions with respect to particular structures.

Thus we see that \textsc{FeynCalc} should not be regarded as a direct competitor to highly automatized packages like e.g. \textsc{FormCalc}, because it neither provides routines for numerical evaluation nor offers a fully automatic workflow to evaluate a scattering process.

For studies that can be carried out using an automatic tool from the beginning to the end, it obviously would not be very efficient to stick to \textsc{FeynCalc}.  While one certainly can chain \textsc{FeynCalc} with appropriate tools and libraries to obtain the same result, this would require more time and effort - which could be invested elsewhere.

There are indeed also other publicly available software packages (\textsc{HEP\-Math} \cite{Wiebusch2014} and \textsc{Package-X} \cite{Patel2015}) that follow the semi-automatic approach to QFT calculations and exhibit many similarities to \textsc{FeynCalc}. Therefore, let us provide a short comparison between those two packages and \textsc{FeynCalc}.

Just as \textsc{FeynCalc}, \textsc{HEPMath} is an open-source project licensed under GPL version 3. This way the users are able to study the source code and possibly modify \textsc{HEPMath} to suit their specific needs. \textsc{Package-X} is on the contrary distributed as a closed source freeware. It can be downloaded directly from the project homepage, but the source code is encrypted which prevents any possible modifications or extensions by the user. Like \textsc{FeynCalc} both \textsc{HEPMath} and  \textsc{Package-X} can manipulate standalone expressions that do not need to represent a valid Feynman diagram.
This is  done by providing special functions for index contractions, treatment of Dirac algebra and manipulations of loop integrals. In $D$-dimensions all three packages can work with anticommuting $\gamma^5$, but only \textsc{HEPMath} and \textsc{FeynCalc} can also treat $\gamma^5$ using the Breitenlohner-Maison-t'Hooft-Veltman scheme \cite{Hooft1972,Breitenlohner1977}. Tensor reduction of 1-loop integrals via Passarino-Veltman technique \cite{Passarino:1978jh} and subsequent simplification of Passarino-Veltman coefficient functions are implemented in each package. \textsc{FeynCalc} 9 and \textsc{Package-X} can work with arbitrarily high-rank tensor integrals, but \textsc{HEPMath} is currently limited to rank 4. As far as the color algebra in concerned, while \textsc{FeynCalc} can deal with general $SU(N)$ generators, HEPMath only supports $SU(3)$ and \textsc{Package-X} does not offer any routines for working with color structures. Both \textsc{FeynCalc} and \textsc{HEPMath} provide an interface to \textsc{FeynArts} for generating Feynman diagrams. \textsc{HEPMath}, however, also contains built-in interfaces to \textsc{LoopTools} \cite{Hahn1999} and \textsc{LHAPDF} \cite{Buckley2015} that are missing in the two other packages. On the other hand, \textsc{Package-X} comes with a library 
of explicit \textit{analytic} results for
1-, 2- and 3-point Passarino-Veltman functions with almost arbitrary kinematics. This is a very useful feature that is, to our knowledge,  not present in any other automatic or semi-automatic packages.

The above comparison shows that all three packages have somewhat different capabilities, but follow essentially the same philosophy to provide the user with convenient and flexible tools for doing calculations in QFT. Since \textsc{HEPMath} and \textsc{Package-X} were released quite recently, they are still not so widely used in research as \textsc{FeynCalc}. However, as future versions of these packages will likely introduce new useful features and improvements, they will also become more visible in the high energy physics community and thus further promote the idea of using semi-automatic tools in suitable computations.

The niche that \textsc{FeynCalc} often fills are calculations that are too specific to be done in a fully automatic fashion but also too challenging to be done (only) by pen and paper, so that semi-automatic evaluation is very welcome.

One example for such problems is the determination of matching coefficients in EFTs. Matching coefficients are extracted by comparing suitable quantities (e.g. Green's functions) between the higher energy theory and its EFT at energies, where both theories should agree by construction. Then the quantity in the higher energy theory usually needs to be expanded in small scales and massaged into a form that resembles the same quantity in the lower energy theory, so that one can read off the values of the matching coefficients.

Such calculations are usually too special to be automatized in a full generality, but they can benefit a lot from functions provided by \textsc{FeynCalc}. This is one of the reasons, why \textsc{FeynCalc} enjoys certain popularity
in the heavy quarkonium physics community, where it is used to perform the matching between QCD and NRQCD for production \cite{ChoPhys.Rev.D53:150-1621996,ZhangPhys.Rev.Lett.98:0920032007,Petrelli1998} or decay \cite{Bodwin2002,Jia2011} of heavy quarkonia. Other studies where \textsc{FeynCalc} was used involve such fields as Higgs \cite{DibJHEP0605:0742006,Buttazzo2013,Florian2013} and top quark physics \cite{Xiao2011}, leptonic decays \cite{Fael2015}
 phenomena in hadronic interactions \cite{Geng2008,Phillips2008}, dark matter \cite{Kopp2009,Cline2010}, neutrino physics \cite{Bray2005,Laha2013,Dasgupta2013} and gravity \cite{Foffa2013}. It is worth noticing that \textsc{FeynCalc} was also used at some stages of NNLO \cite{Gaunt2014,Feng2015} calculations. Indeed, \textsc{FeynCalc} can be well employed for small or medium size multi-loop processes if one connects it to suitable tools for IBP-reduction (e.g. \textsc{FIRE} \cite{Smirnov2013}) and numeric evaluation of multi-loop integrals (e.g. FIESTA \cite{Smirnov2013a} or \textsc{SecDec} \cite{Borowka2012}).

Last but not least, \textsc{FeynCalc} can be also useful for educational purposes.
The possibility of easily getting hands-on experience with computing Feynman diagrams and exploring the different steps involved can be very
helpful and motivating for students of quantum field theory.

\section{New features in \textsc{FeynCalc} 9.0}

\subsection{Improved tensor decomposition} \label{sec:newf}

In the very early versions of \textsc{FeynCalc}, tensor decomposition of 1-loop integrals (via Passarino-Veltman technique \cite{Passarino:1978jh}) could be done only using the function \texttt{OneLoop}, where the maximal rank of the integrals was limited to 4 and the output was always written in terms of Passarino-Veltman coefficient functions. Although one could reduce Passarino-Veltman coefficient functions with rank higher than 4 using \texttt{PaVeReduce}, the
tensor basis for such higher rank integrals had to be constructed by hand.

While working on \cite{Mertig1996}, {Rolf} {Mertig} added to  \textsc{FeynCalc} 3.0 a tool (\texttt{Tdec}) for tensor decomposition of multi-loop integrals of arbitrary rank and multiplicity (for non-zero Gram determinants) and even included a database (\texttt{TIDL}) to load already computed decompositions, but only a very small amount of this functionality was turned into a user-friendly routine \texttt{TID} (1-loop only), while the rest remained to ``lie idle'' in the source code. \texttt{TID} was limited to 4-point functions of rank 4 and could not handle kinematic configurations with zero Gram determinants, so that for such cases one was forced to use \texttt{OneLoop}. However, in \textsc{FeynCalc} 4 the reduction of rank 4 tensor integrals via \texttt{OneLoop} was disabled due to its poor efficiency. 
 As a consequence of all these developments the tensor reduction of 1-loop integrals (especially with rank higher than 3) in the recent \textsc{FeynCalc} versions often turned to be cumbersome and inconvenient.

In \textsc{FeynCalc} 9.0 \texttt{TID} was rewritten almost from scratch to allow for 1-loop tensor decompositions of any rank and multiplicity. At the beginning, the function computes Gram determinants for all the unique 1-loop integrals in the expression. If the determinant vanishes, the decomposition for that integral is done in terms of the Passarino-Veltman coefficient functions.

\begin{lstlisting}[language=Mathematica,mathescape,escapechar=!]

In[1]:= FCClearScalarProducts[];
        ScalarProduct[p1, p1] = 0;
        int = FCI[SPD[p2, q] FAD[{q, m0}, {q + p1, m1}]]

Out[1]:= $\frac{\text{p2}\cdot q}{\left(q^2-\text{m0}^2\right).\left((\text{p1}+q)^2-\text{m1}^2\right)}$

In[2]:= TID[int, q]

Out[2]:= $i \pi ^2 (\text{p1}\cdot \text{p2}) \text{B}_1\left(0,\text{m0}^2,\text{m1}^2\right)$

\end{lstlisting}
Otherwise, \texttt{TID} will output the result in terms of scalar 1-loop integrals.

\begin{lstlisting}[language=Mathematica,mathescape,escapechar=!]

In[1]:= FCClearScalarProducts[];
        int = FCI[SPD[p2, q] FAD[{q, m0}, {q + p1, m1}]]

Out[1]:= $-\frac{\left(\text{m0}^2-\text{m1}^2+\text{p1}^2\right) (\text{p1}\cdot \text{p2})}{2 \text{p1}^2 \left(q^2-\text{m0}^2\right).\left((q-\text{p1})^2-\text{m1}^2\right)}+\frac{\text{p1}\cdot \text{p2}}{2 \text{p1}^2 \left(q^2-\text{m0}^2\right)}-\frac{\text{p1}\cdot \text{p2}}{2 \text{p1}^2 \left(q^2-\text{m1}^2\right)}$

\end{lstlisting}

If needed, those scalar integrals can be converted to Passarino-Veltman scalar functions by using \texttt{ToPaVe}, which is also available since \textsc{FeynCalc} 9.0.

\begin{lstlisting}[language=Mathematica,mathescape,escapechar=!]

In[2]:= TID[int, q] // ToPaVe[#, q] &

Out[2]:= $-\frac{i \pi ^2 \left(\text{m0}^2-\text{m1}^2+\text{p1}^2\right) (\text{p1}\cdot \text{p2}) \text{B}_0\left(\text{p1}^2,\text{m0}^2,\text{m1}^2\right)}{2 \text{p1}^2}+\frac{i \pi ^2 \text{A}_0\left(\text{m0}^2\right) (\text{p1}\cdot \text{p2})}{2 \text{p1}^2}-\frac{i \pi ^2 \text{A}_0\left(\text{m1}^2\right) (\text{p1}\cdot \text{p2})}{2 \text{p1}^2}$
\end{lstlisting}

The decompositions in terms of scalar integrals tend to become very large already for 3-point functions, so to obtain more compact expressions it might be desirable to use the basis of Passarino-Veltman coefficient functions, even if there are no zero Gram determinants. This can be easily achieved via the option \texttt{UsePaVeBasis}.

\begin{lstlisting}[language=Mathematica,mathescape,escapechar=!]

In[1]:= int = FCI[FVD[q, mu] FVD[q, nu] FAD[{q, m0}, {q + p1, m1}, {q + p2, m2}]]

Out[1]:= $\frac{q^{\text{mu}} q^{\text{nu}}}{\left(q^2-\text{m0}^2\right).\left((\text{p1}+q)^2-\text{m1}^2\right).\left((\text{p2}+q)^2-\text{m2}^2\right)}$

In[2]:= TID[int/(I*Pi^2), q, UsePaVeBasis -> True]

Out[2]:=  $g^{\text{mu}\text{nu}} \text{C}_{00}\left(\text{p1}^2,-2 (\text{p1}\cdot   \text{p2})+\text{p1}^2+\text{p2}^2,\text{p2}^2,\text{m0}^2,\text{m1}^2,\text{m2}^2\right)$
          $+\text{p1}^{\text{mu}} \text{p1}^{\text{nu}} \text{C}_{11}\left(\text{p1}^2,-2 (\text{p1}\cdot \text{p2})+\text{p1}^2+\text{p2}^2,\text{p2}^2,\text{m0}^2,\text{m1}^2,\text{m2}^2\right)$
          $+\left(\text{p2}^{\text{mu}} \text{p1}^{\text{nu}} +\text{p1}^{\text{mu}} \text{p2}^{\text{nu}}\right) \text{C}_{12}\left(\text{p1}^2,-2 (\text{p1}\cdot \text{p2})+\text{p1}^2+\text{p2}^2,\text{p2}^2,\text{m0}^2,\text{m1}^2,\text{m2}^2\right)$
         $+\text{p2}^{\text{mu}} \text{p2}^{\text{nu}} \text{C}_{22}\left(\text{p1}^2,-2 (\text{p1}\cdot \text{p2})+\text{p1}^2+\text{p2}^2,\text{p2}^2,\text{m0}^2,\text{m1}^2,\text{m2}^2\right)$

\end{lstlisting}

All the Passarino-Veltman functions are defined as in \textsc{LoopTools} \cite{Hahn1999} and explicit definitions are encoded for functions with up to 5 legs. For integrals with even higher multiplicities  the coefficient functions (denoted as \texttt{GenPaVe}) simply include the dependence on the external momenta that can be used to convert them to the \textsc{LoopTools} or any other convention.

\begin{lstlisting}[language=Mathematica,mathescape,escapechar=!]

In[1]:= int = FCI[FVD[q, mu] FVD[q,nu] FAD[{q, m0}, {q, m1}, {q, m2}, {q, m3}, {q +       p4, m4}, {q + p5, m5}, {q + p6, m6}]]

Out[1]:= $\left(q^{\text{mu}} q^{\text{nu}}\right)/\left(q^2-\text{m0}^2\right).\left(q^2-\text{m1}^2\right).\left((\text{p2}+q)^2-\text{m2}^2\right).\left((\text{p3}+q)^2-\text{m3}^2\right).$
         $\left((\text{p4}+q)^2-\text{m4}^2\right).\left((\text{p5}+q)^2-\text{m5}^2\right).\left((\text{p6}+q)^2-\text{m6}^2\right)$

In[2]:= TID[int/(I*Pi^2), q, UsePaVeBasis -> True]

Out[2]:=
$g^{\text{mu}\text{nu}} \text{GenPaVe}\left(\{0,0\},\left(
\begin{array}{cc}
 0 & \text{m0} \\
 \text{p1} & \text{m1} \\
 \text{p2} & \text{m2} \\
 \text{p3} & \text{m3} \\
 \text{p4} & \text{m4} \\
 \text{p5} & \text{m5} \\
 \text{p6} & \text{m6}
\end{array}
\right)\right)+\text{p1}^{\text{mu}} \text{p1}^{\text{nu}} \text{GenPaVe}\left(\{1,1\},\left(
\begin{array}{cc}
 0 & \text{m0} \\
 \text{p1} & \text{m1} \\
 \text{p2} & \text{m2} \\
 \text{p3} & \text{m3} \\
 \text{p4} & \text{m4} \\
 \text{p5} & \text{m5} \\
 \text{p6} & \text{m6}
\end{array}
\right)\right)  +  \ldots$
\end{lstlisting}
Here, \texttt{GenPaVe[\{1,1\},\{\{0,m0\},\{Momentum[p1],m1\}, \ldots, \\ \{Momentum[p6],m6\}\}]} stands for the coefficient function of $p_1^\mu p_1^\nu$ in the tensor decomposition of

\begin{equation}
\int d^D q \, \frac{q^\mu q^\nu}{[q^2-m_0^2][(q-p_1)^2-m_1^2][(q-p_2)^2-m_2^2]\cdots [(q-p_6)^2-m_6^2]}.
\end{equation}

Since this kind of output is useful if one explicitly wants  to obtain coefficient functions defined in a different way than in \textsc{LoopTools}, it can be activated also for functions with lower multiplicities by setting the option \texttt{GenPaVe} of \texttt{TID} to \texttt{True}.

\begin{lstlisting}[language=Mathematica,mathescape,escapechar=!]
In[1]:= int = FCI[FVD[q, mu] FVD[q, nu] FAD[{q, m0}, {q + p1, m1}]]

Out[1]:= $\frac{q^{\text{mu}} q^{\text{nu}}}{\left(q^2-\text{m0}^2\right).\left((\text{p1}+q)^2-\text{m1}^2\right)}$

In[2]:= TID[int/(I*Pi^2), q, UsePaVeBasis -> True, GenPaVe -> True]

Out[2]:=
      $g^{\text{mu}\text{nu}} \text{GenPaVe}\left(\{0,0\},\left(
\begin{array}{cc}
 0 & \text{m0} \\
 \text{p1} & \text{m1}
\end{array}
\right)\right)+\text{p1}^{\text{mu}} \text{p1}^{\text{nu}} \text{GenPaVe}\left(\{1,1\},\left(
\begin{array}{cc}
 0 & \text{m0} \\
 \text{p1} & \text{m1}
\end{array}
\right)\right)$
\end{lstlisting}

One should also keep in mind that \textsc{FeynCalc} cannot perform any further simplifications of \texttt{GenPaVe} functions, because internally they are not  recognized as Passarino-Veltman integrals (\texttt{PaVe}).

It is well known that for a general multi-loop multi-scale integral, tensor decomposition does not allow to cancel all the scalar products containing loop momenta in the numerator, as it is the case at 1-loop. Nevertheless, this technique is widely used also in calculations beyond 1-loop, especially if one needs to deal with integrals that have loop momenta contracted to Dirac matrices or epsilon tensors or even loop momenta with free Lorentz indices. \textsc{FeynCalc} uses a special reduction algorithm (implemented in \texttt{Tdec}) that consists of decomposing the integral into all tensor structures allowed by the symmetries and using a modified version of Gaussian elimination to obtain the coefficients of each tensor.

 Since tensor decomposition of multi-loop integrals with \textsc{FeynCalc}'s function \texttt{Tdec} is not very straight-forward and usually requires some additional \textsc{Mathematica} code, in \textsc{FeynCalc} 9.0 a new function \texttt{FCMultiLoop\-TID} was added, that makes multi-loop tensor reduction work out of the box.

\begin{lstlisting}[language=Mathematica,mathescape,escapechar=!]
In[1]:= int = FCI[FVD[q1, mu] FVD[q2, nu] FAD[q1, q2, {q1 - p1},
        {q2 - p1}, {q1 - q2}]]

Out[1]:= $\frac{\text{q1}^{\text{mu}} \text{q2}^{\text{nu}}}{\text{q1}^2.\text{q2}^2.(\text{q1}-\text{p1})^2.(\text{q2}-\text{p2})^2.(\text{q1}-\text{q2})^2}$

In[2]:= FCMultiLoopTID[int, {q1, q2}]

Out[2]:=   $\frac{D \text{p1}^{\text{mu}} \text{p1}^{\text{nu}}-\text{p1}^2 g^{\text{mu}\text{nu}}}{4 (D-1) \text{q2}^2.\text{q1}^2.(\text{q2}-\text{p1})^2.(\text{q1}-\text{q2})^2.(\text{q1}-\text{p1})^2}-\frac{D \text{p1}^{\text{mu}} \text{p1}^{\text{nu}}-\text{p1}^2 g^{\text{mu}\text{nu}}}{2 (D-1) \text{p1}^4 \text{q1}^2.(\text{q2}-\text{p1})^2.(\text{q1}-\text{q2})^2}$

             $+\frac{\text{p1}^2 g^{\text{mu}\text{nu}}-\text{p1}^{\text{mu}} \text{p1}^{\text{nu}}}{(D-1) \text{p1}^2 \text{q2}^2.\text{q1}^2.(\text{q1}-\text{q2})^2.(\text{q1}-\text{p1})^2}-\frac{\text{p1}^2 g^{\text{mu}\text{nu}}-\text{p1}^{\text{mu}} \text{p1}^{\text{nu}}}{2 (D-1) \text{p1}^2 \text{q2}^2.\text{q1}^2.(\text{q2}-\text{p1})^2.(\text{q1}-\text{p1})^2}$
\end{lstlisting}

Unfortunately, the reduction breaks down when the corresponding Gram determinant vanishes. For such cases, in a future version it is planned to include a more useful algorithm.

\subsection{New partial fractioning algorithm}

Since the version 3, \textsc{Feyn\-Calc} includes \texttt{Scalar\-Product\-Cancel} and \\ \texttt{Apart2} that can be used to rewrite loop integrals in a simpler form.  \texttt{Scalar\-Product\-Cancel} essentially applies the well known identity \cite{Passarino:1978jh}

\begin{equation}
q \cdot p = \frac{1}{2} [(q+p)^2 +m_2^2 -(q^2 + m_1^2) - p^2 -m_2^2 + m_1^2]
\end{equation}
repeatedly, until all scalar products containing loop momenta that can be canceled in this way are eliminated. \texttt{Apart2} uses the trivial identity

\begin{equation}
\frac{1}{(q^2-m_1^2)(q^2-m_2^2)} = \frac{1}{m_1^2-m_2^2} \left ( \frac{1}{q^2-m_1^2} -\frac{1}{q^2-m_2^2} \right )
\end{equation}
to simplify suitable denominators. In principle, these two functions implement some aspects of partial fractioning, i.e., the decomposition of a loop integral with linearly dependent propagators into a sum of integrals where each integral contains only linearly independent propagators.
Notice that here we count scalar products that involve loop momenta as propagators with negative exponents. Unfortunately, there are plenty of examples where neither \texttt{ScalarProductCancel} nor \texttt{Apart2} can partial fraction an integral with linearly dependent propagators, e.g.

\begin{equation}
\int d^D q \, \frac{1}{q^2 (q-p)^2 (q+p)^2} \\
= \frac{1}{p^2} \int d^D q \left(  \frac{1}{q^2(q-p)^2} - \frac{1}{(q-p)^2(q+p)^2} \right)
\end{equation}

A general partial fractioning algorithm that is suitable for multi-loop integrals including its \textsc{Mathematica}  implementation (\textsc{APart}\footnote{\url{https://github.com/F-Feng/APart}})  was presented in \cite{Feng2012a}. The author has also shown how his code can be used together with \textsc{FeynCalc} in order to decompose different loop integrals. For this the user is required to convert a loop integral in the \textsc{FeynCalc} notation (with denominator encoded in \texttt{FeynAmp\-Denominator}) to a somewhat different form and to specify all the scalar products that contain loop momenta and appear in this loop integral. After the decomposition the resulting integrals need to be converted back into \textsc{FeynCalc} notation.

In \textsc{FeynCalc} 9.0 the algorithm from \cite{Feng2012a}  was adopted and reimplemented to be the standard partial fractioning routine. As such, it is fully integrated with all other \textsc{FeynCalc} functions and objects and does not require any explicit conversion of the input or output.

\begin{lstlisting}[language=Mathematica,mathescape,escapechar=!]
In[1]:= int = FAD[{q}, {q - p}, {q + p}]

Out[1]:= $\frac{1}{q^2.(q-p)^2.(p+q)^2}$

In[2]:= ApartFF[int1, {q}]

Out[2]:=   $\frac{1}{p^2 q^2.(q-p)^2}-\frac{1}{p^2 q^2.(q-2 p)^2}$
\end{lstlisting}

The name of the corresponding function is \texttt{ApartFF} which stands for ``Apart Feng Feng'' and serves as an additional acknowledgement of the original author. One should also notice that while the original \textsc{APart} can be used for partial fractioning of general multivariate polynomials, the \textsc{FeynCalc} version is limited only to polynomials that appear in Feynman diagrams as propagators and scalar products. Thus, it is much less general than \textsc{APart} but is also more convenient when used with \textsc{FeynCalc}.

\subsection{Tools for interfacing \textsc{FeynCalc} with packages for IBP-reduction}

In modern multi-loop calculations, reduction of scalar loop integrals via integration-by-parts (IBP) identities \cite{Chetyrkin1981} is a regular step needed to arrive to a smaller set of master integrals.

Although \textsc{FeynCalc} does not include a general purpose tool for IBP reduction (the built-in TARCER \cite{Mertig1998} is suitable only for 2-loop self-energy type integrals), this omission can be compensated by using one of the publicly available IBP-packages (\textsc{FIRE} \cite{Smirnov2013},
\textsc{LiteRED} \cite{Lee2012},
\textsc{Reduze} \cite{Studerus2009},
\textsc{AIR} \cite{Anastasiou2004}). However, one should keep in mind that such tools usually expect their input to contain only loop integrals with linearly independent propagators that form a basis.
For example, the integral

\begin{equation}
\int d^D q_1 \, d^D q_2 \, d^D q_3 \, \frac{1}{[q_1^2-m^2]^2[(q_1+q_3)^2-m^2](q_2-q_3)^2 q_2^2}
\end{equation}
cannot be processed by \textsc{FIRE} in this form because $q_1^2$, $q_2^2$, $(q_1+q_3)^2$ and $(q_2-q_3)^2$ alone do not form a basis.

\begin{lstlisting}[language=Mathematica,mathescape,escapechar=!]
In[1]:= << FIRE5`FIRE5`
        Internal = {q1, q2, q3};
        External = {};
        Propagators = {q1^2 - m^2, (q1 + q3)^2 - m^2, (q2 - q3)^2, q2^2};
        PrepareIBP[];

Out[1]:= FIRE, version 5.1
         DatabaseUsage: 0
         UsingFermat: False
         Not enough propagators. Add irreducible nominators
\end{lstlisting}

If one includes also $q_3^2$ and $q_1 \cdot q_2$ with zero exponentials, then we have a proper basis and the reduction works as it should. Also the integral
\begin{equation}
\int d^D q_1 \, d^D q_2 \, \frac{(p \cdot q_1)^2 (p \cdot q_2)^2}{[q_1^2-m^2] [q_2^2-m^2] (q_1-p)^2(q_2-p)^2 (q_1-q_2)^2}
\end{equation}
cannot be reduced right away, this time because its propagators are linearly dependent.

\begin{lstlisting}[language=Mathematica,mathescape,escapechar=!]
In[1]:= << FIRE5`FIRE5`
        Internal = {q1, q2};
        External = {};
        Propagators = {q1^2 - m^2, q2^2 - m^2, (q1 - p)^2, (q2 - p)^2, (q1 - q2)^2, p q1, p q2};
        PrepareIBP[];

Out[1]:= FIRE, version 5.1
         DatabaseUsage: 0
         UsingFermat: False
         Linearly dependant propagators. Perform reduction first
\end{lstlisting}

To detect such problems before the reduction actually fails, \textsc{FeynCalc} 9.0 introduces two new special functions. When \texttt{FCLoopBasisIncompleteQ} is applied to a loop integral, it returns \texttt{True} if this integral does not contain enough irreducible propagators.
\begin{lstlisting}[language=Mathematica,mathescape,escapechar=!]
In[1]:= intP1 = FCI[FAD[{q1, m, 2}, {q1 + q3, m}, {q2 - q3}, q2]]

Out[1]:= $\frac{1}{\left(\text{q1}^2-m^2\right).\left(\text{q1}^2-m^2\right).\left((\text{q1}+\text{q3})^2-m^2\right).(\text{q2}-\text{q3})^2.\text{q2}^2}$

In[2]:= FCLoopBasisIncompleteQ[intP1, {q1, q2, q3}]
Out[2]:= True

In[3]:= FCLoopBasisIncompleteQ[SPD[q3, q3] SPD[q1, q2] intP1, {q1, q2, q3}]
Out[3]:= False
\end{lstlisting}

An integral with linearly dependent propagators will be detected by \,
\texttt{FCLoopBasisOverdeterminedQ},
\begin{lstlisting}[language=Mathematica,mathescape,escapechar=!]
In[1]:= intP2 = FCI[SPD[p, q1]^2 SPD[p, q2]^2 FAD[{q1, m}, {q2, m}, q1 - p, q2 - p, q1 - q2]]

Out[1]:= $\frac{(p\cdot \text{q1})^2 (p\cdot \text{q2})^2}{\left(\text{q1}^2-m^2\right).\left(\text{q2}^2-m^2\right).(\text{q1}-p)^2.(\text{q2}-p)^2.(\text{q1}-\text{q2})^2}$

In[2]:= FCLoopBasisOverdeterminedQ[intP2, {q1, q2, q3}]
Out[2]:= True

\end{lstlisting}
so that only an integral for which both functions return \texttt{False} can be reduced in a straight-forward way.

In a practical calculation where one knows what integral topologies are involved, such issues can be easily resolved. In particular, a clever choice of additional propagators that are needed to have a basis, can greatly simplify the reduction. On the other hand, depending on the size of the problem and the number of topologies involved, a less clever but fully automatic solution may also be useful.

For an integral with linearly dependent propagators we can use \texttt{ApartFF}, that is guaranteed to decompose it into integrals where all propagators are linearly independent.

\begin{lstlisting}[language=Mathematica,mathescape,escapechar=!]
In[1]:= ApartFF[intP2, {q1, q2}]

Out[1]:= $\frac{\left(m^2+p^2\right)^4}{16 \left(\text{q1}^2-m^2\right).\left(\text{q2}^2-m^2\right).(\text{q2}-p)^2.(\text{q1}-\text{q2})^2.(\text{q1}-p)^2}$
          $-\frac{\left(m^2+p^2\right)^3}{8 \left(\text{q1}^2-m^2\right).\left(\text{q2}^2-m^2\right).(\text{q1}-\text{q2})^2.(\text{q1}-p)^2}$
          $-\frac{\left(m^2+p^2\right) (p\cdot \text{q1})}{8 \left(\text{q2}^2-m^2\right).(\text{q1}-\text{q2})^2.(\text{q1}-p)^2} + \ldots$

\end{lstlisting}

For integrals with an incomplete basis of propagators one can use the new function \texttt{FCLoopBasisFindCompletion} that finds out which irreducible propagators (with zero exponents) are missing.

\begin{lstlisting}[language=Mathematica,mathescape,escapechar=!]
In[1]:= FCLoopBasisFindCompletion[intP1, {q1, q2, q3}]

Out[1]:= $\left\{\frac{1}{\left(\text{q1}^2-m^2\right).\left(\text{q1}^2-m^2\right).\left((\text{q1}+\text{q3})^2-m^2\right).(\text{q2}-\text{q3})^2.\text{q2}^2},\left\{-(\text{q1}\cdot \text{q3})+\text{q2}\cdot \text{q3}+2 \text{q3}^2,\text{q1}\cdot \text{q2}\right\}\right\}$

\end{lstlisting}
With the suggested propagators the integral is guaranteed to have a complete basis, but the choice of the propagators themselves is usually not very clever.
This is because in general \textsc{FeynCalc} cannot guess the topology of the given integral without any additional input. It is planned to provide a possibility for specifying the topology, which would admittedly make
\texttt{FCLoopBasisFindCompletion} much more useful than it is now.

Still, with \texttt{ApartFF} and \texttt{FCLoopBasisFindCompletion} it is now possible to automatically bring any scalar multi-loop integral in \textsc{FeynCalc} notation to a form that can be directly (modulo notation conversion) forwarded to an IBP tool.

\subsection{Advanced extraction of loop integrals}

The idea to use \textsc{FeynCalc} as a sort of switch board for different computational tools in a larger framework (see e.g. \cite{Feng2013}) is further developed in version 9.0 by the introduction of new functions that can extract different loop integrals from the given expression.

One of them, \texttt{FCLoopSplit}, breaks the given expression into four pieces, which are

\begin{enumerate}
\item Terms that contain no loop integrals.
\item Terms that only contain scalar loop integrals without any loop momenta in the  denominators, e.g.
\begin{equation}
\int d^D q \, \frac{1}{q^2-m^2}.
\end{equation}
\item Terms that contain scalar loop integrals with loop momenta dependent scalar products in the denominators, e.g.
\begin{equation}
\int d^D q \, \frac{(q \cdot p)}{q^2 (q-p)^2}.
\end{equation}
\item Terms that contain tensor loop integrals, e.g.
\begin{equation}
\int d^D q \, \frac{q^\mu q^\nu}{q^2-m^2} \quad  \textrm{or} \quad \int d^D q  \,\frac{ (\gamma \cdot q)}{q^2 (q-p)^2}.
\end{equation}
\end{enumerate}

\begin{lstlisting}[language=Mathematica,mathescape,escapechar=!]
In[1]:= int = FCI[(GSD[q - p] + m).GSD[x] FAD[q, {q - p, m}] + (m^2 + SPD[q, q]) FAD[{q, m,2}]];
Out[1]:= $\frac{(m+\gamma \cdot (q-p)).(\gamma \cdot x)}{q^2.\left((q-p)^2-m^2\right)}+\frac{m^2+q^2}{\left(q^2-m^2\right).\left(q^2-m^2\right)}$

In[2]:= FCLoopSplit[int, {q}]

Out[2]:= $\left\{0,\frac{m \gamma \cdot x-(\gamma \cdot p).(\gamma \cdot x)}{q^2.\left((q-p)^2-m^2\right)}+\frac{m^2}{\left(q^2-m^2\right).\left(q^2-m^2\right)},\frac{q^2}{\left(q^2-m^2\right).\left(q^2-m^2\right)},\frac{(\gamma \cdot q).(\gamma \cdot x)}{q^2.\left((q-p)^2-m^2\right)}\right\}$

\end{lstlisting}

This splitting makes it easier to handle different types of loop integrals and to simplify them with \textsc{FeynCalc} or other tools. For example, if one wants to perform tensor reduction of multi-loop integrals with \textsc{FaRe} \cite{Fiorentin2015} instead of \texttt{FCMultiLoopTID}, it can be done by applying \texttt{FCLoopSplit} to the given expression and working with the fourth element of the resulting list, while the other elements remain unchanged and can be later added to the final expression.

To handle a larger number of loop diagrams  in an efficient way, \texttt{FCLoop\-Split} alone is not sufficient. This is because same integrals may appear multiple times in different diagrams and ignoring this fact would make the evaluation more complex than it actually is. To avoid this kind of problems one should better first analyze the amplitude and extract all the unique integrals. Then each unique integral needs to be evaluated only once, no matter how often it appears in the full expression.
In \textsc{FeynCalc} 9.0 this can be conveniently done with
\texttt{FCLoopIsolate}. The function wraps loop integers with the given head, such that the list of unique integrals can be quickly created with \textsc{Mathematica}'s \texttt{Cases} and \texttt{Union} or just \textsc{FeynCalc}'s \texttt{Cases2}

\begin{lstlisting}[language=Mathematica,mathescape,escapechar=!]
In[1]:= int = FCI[GSD[q - p1].(GSD[q - p2] + M).GSD[p3] SPD[q, p2] FAD[q, q - p1, {q - p2, m}]];

Out[1]:= $\frac{(\text{p2}\cdot q) (\gamma \cdot (q-\text{p1})).(M+\gamma \cdot (q-\text{p2})).(\gamma \cdot \text{p3})}{q^2.(q-\text{p1})^2.\left((q-\text{p2})^2-m^2\right)}$

In[2]:= res = FCLoopIsolate[int, {q}, Head -> loopInt]

Out[2]:= $\text{loopInt}\left(\frac{\text{p2}\cdot q}{q^2.(q-\text{p1})^2.\left((q-\text{p2})^2-m^2\right)}\right) ((\gamma \cdot \text{p1}).(\gamma \cdot \text{p2}).(\gamma \cdot \text{p3})-M (\gamma \cdot \text{p1}).(\gamma \cdot \text{p3}))+$

     $M \text{loopInt}\left(\frac{(\text{p2}\cdot q) (\gamma \cdot q).(\gamma \cdot \text{p3})}{q^2.(q-\text{p1})^2.\left((q-\text{p2})^2-m^2\right)}\right)-\text{loopInt}\left(\frac{(\text{p2}\cdot q) (\gamma \cdot \text{p1}).(\gamma \cdot q).(\gamma \cdot \text{p3})}{q^2.(q-\text{p1})^2.\left((q-\text{p2})^2-m^2\right)}\right)$

     $-\text{loopInt}\left(\frac{(\text{p2}\cdot q) (\gamma \cdot q).(\gamma \cdot \text{p2}).(\gamma \cdot \text{p3})}{q^2.(q-\text{p1})^2.\left((q-\text{p2})^2-m^2\right)}\right)+\text{loopInt}\left(\frac{(\text{p2}\cdot q) (\gamma \cdot q).(\gamma \cdot q).(\gamma \cdot \text{p3})}{q^2.(q-\text{p1})^2.\left((q-\text{p2})^2-m^2\right)}\right)$

In[3]:= Cases2[res, loopInt]

Out[3]:= $\left\{\text{loopInt}\left(\frac{\text{p2}\cdot q}{q^2.(q-\text{p1})^2.\left((q-\text{p2})^2-m^2\right)}\right),\text{loopInt}\left(\frac{(\text{p2}\cdot q) (\gamma \cdot q).(\gamma \cdot \text{p3})}{q^2.(q-\text{p1})^2.\left((q-\text{p2})^2-m^2\right)}\right),\right.$

          $\left. \text{loopInt}\left(\frac{(\text{p2}\cdot q) (\gamma \cdot \text{p1}).(\gamma \cdot q).(\gamma \cdot \text{p3})}{q^2.(q-\text{p1})^2.\left((q-\text{p2})^2-m^2\right)}\right), \text{loopInt}\left(\frac{(\text{p2}\cdot q) (\gamma \cdot q).(\gamma \cdot \text{p2}).(\gamma \cdot \text{p3})}{q^2.(q-\text{p1})^2.\left((q-\text{p2})^2-m^2\right)}\right),\right. $

          $\left. \text{loopInt}\left(\frac{(\text{p2}\cdot q) (\gamma \cdot q).(\gamma \cdot q).(\gamma \cdot \text{p3})}{q^2.(q-\text{p1})^2.\left((q-\text{p2})^2-m^2\right)}\right)\right\}$

\end{lstlisting}

The combined application of \texttt{FCLoopIsolate} and \texttt{FCLoopSplit} is provided by \texttt{FCLoopExtract}. This function returns a list of three entries. The first one contains the part of the expression which is free of loop integrals. The second entry consists of the remaining expression where every loop integral is wrapped with the given head. Finally, the last entry contains a list of all the unique loop integrals in the expression.

\begin{lstlisting}[language=Mathematica,mathescape,escapechar=!]
In[4]:= FCLoopExtract[int, {q}, loopInt][[1]]

Out[4]:= 0

In[5]:= FCLoopExtract[int, {q}, loopInt][[2]] ===
 FCLoopIsolate[int, {q}, Head -> loopInt]
 
Out[5]:= True

In[6]:= FCLoopExtract[int, {q}, loopInt][[3]] ===
 Cases2[res, loopInt]
 
Out[6]:= True

\end{lstlisting}

Suppose that we want to evaluate these loop integrals using some custom function \texttt{loopEval} (in this example it is just a dummy function that computes the hash of each loop integral). All we need to do is to apply \texttt{FCLoopExtract} to the initial expression, map the list of the unique integrals to \texttt{loopEval}, create a substitution rule and apply this rule to our expression in order to get the final result.

\begin{lstlisting}[language=Mathematica,mathescape,escapechar=!]
In[7]:= {rest, loops, intsUnique} = FCLoopExtract[int, {q}, loopInt];

In[8]:= loopEval[x_] := ToString[Hash[x]];

In[9]:= solsList = loopEval /@ uniqueInts

Out[9]:= {2069116068,115167616,776830638,1878762839,1337833147}

In[10]:= repRule = MapThread[Rule[#1, #2] &, {intsUnique, solsList}]
Out[10]:= $\left\{\text{loopInt}\left(\frac{\text{p2}\cdot q}{q^2.(q-\text{p1})^2.\left((q-\text{p2})^2-m^2\right)}\right)\to 2069116068,\right.$

          $\left.\text{loopInt}\left(\frac{(\text{p2}\cdot q) (\gamma \cdot q).(\gamma \cdot \text{p3})}{q^2.(q-\text{p1})^2.\left((q-\text{p2})^2-m^2\right)}\right)\to 115167616,\right.$

          $\left.\text{loopInt}\left(\frac{(\text{p2}\cdot q) (\gamma \cdot \text{p1}).(\gamma \cdot q).(\gamma \cdot \text{p3})}{q^2.(q-\text{p1})^2.\left((q-\text{p2})^2-m^2\right)}\right)\to 776830638,\right.$
          $\left.\text{loopInt}\left(\frac{(\text{p2}\cdot q) (\gamma \cdot q).(\gamma \cdot \text{p2}).(\gamma \cdot \text{p3})}{q^2.(q-\text{p1})^2.\left((q-\text{p2})^2-m^2\right)}\right)\to 1878762839,\right.$
          $\left.\text{loopInt}\left(\frac{(\text{p2}\cdot q) (\gamma \cdot q).(\gamma \cdot q).(\gamma \cdot \text{p3})}{q^2.(q-\text{p1})^2.\left((q-\text{p2})^2-m^2\right)}\right)\to 1337833147\right\}$

Int[11]:= res = rest + loops /. repRule

Out[11]:= $115167616 M+1337833147-1878762839$+

           $2069116068 ((\gamma \cdot \text{p1}).(\gamma \cdot \text{p2}).(\gamma \cdot \text{p3})- M (\gamma \cdot \text{p1}).(\gamma \cdot \text{p3}))-776830638$

\end{lstlisting}

With \texttt{FCLoopSplit}, \texttt{FCLoopIsolate} and \texttt{FCLoopExtract} it is now much easier not only to manipulate loop integrals, but also to check which integrals actually appear in an expression. Unique loop integrals can be evaluated with tools outside of \textsc{FeynCalc} and then substituted back by just a couple of lines of \textsc{Mathematica} code.

\subsection{Better interface to \textsc{FeynArts}}

If \textsc{FeynCalc} needs to be used with a Feynman diagram generator, then \textsc{FeynArts} is usually the most convenient choice. Initially the syntax of both packages was adjusted to make them fully compatible with each other. In fact, for the very first version of \textsc{FeynArts} \cite{Kueblbeck1990}, \textsc{FeynCalc} was referred to as the standard tool to evaluate the generated amplitudes.
As \textsc{FeynArts} was developed further, the full compatibility was lost, but even now, the output of \textsc{FeynArts} can be converted into valid \textsc{FeynCalc} input with only little effort. A more severe problem in using this setup arises when \textsc{FeynArts} and \textsc{FeynCalc} are loaded in the same \textsc{Mathematica} session. Unfortunately, both packages contain objects with same names but different contexts, definitions and properties (e.g. \texttt{FourVector}, \texttt{DiracMatrix} or \texttt{FeynAmpDenominator}) such that it is not possible to use them together without risking inconsistencies. To avoid these issues \textsc{FeynCalc} is able to automatically patch the source code of \textsc{FeynArts} by renaming all the conflicting symbols, such that e.g. \texttt{FourVector} becomes \texttt{FAFourVector} and no variable shadowing can occur. This patching mechanism was greatly improved in \textsc{FeynCalc} 9.0 both in terms of user friendliness and compatibility to other \textsc{Mathematica} packages. The patched copy of \textsc{FeynArts} now resides in the \textit{FeynArts} directory inside the \textsc{FeynCalc} installation. By default this directory is empty. The user is expected to manually download the latest \textsc{FeynArts} tarball from the official website\footnote{http://www.feynarts.de}  and unpack its content to \textit{FeynCalc/FeynArts}. When \textsc{FeynCalc} is loaded via

\begin{lstlisting}[language=Mathematica,mathescape,escapechar=!]

$\$$LoadFeynArts=True;
<<FeynCalc`
\end{lstlisting}
it will automatically detect \textsc{FeynArts} installation and offer the user to patch it. This procedure is required only once and after that one can use \textsc{FeynArts} and \textsc{FeynCalc} together without any problems.

After all the required diagrams have been generated and turned into amplitudes with \textsc{FeynArts}` function \texttt{CreateFeynAmp}, the output still needs to be converted into valid \textsc{FeynCalc} input. In \textsc{FeynCalc} 9.0 this is handled by the new function \texttt{FCFAConvert} that takes the output of \texttt{CreateFeynAmp} and generates proper \textsc{FeynCalc} expressions based on the given options.  With \texttt{IncomingMomenta}, \texttt{OutgoingMomenta} and \texttt{LoopMomenta} the user can specify how the corresponding momenta should be named. Otherwise they will be denoted as \texttt{InMom1}, \texttt{InMom2}, \ldots , \texttt{OutMom1}, \texttt{OutMom2}, \ldots and \texttt{LoopMom1}, \texttt{LoopMom2}, \ldots. Polarization vectors of external massless bosons are by default not transverse, but can be made so if the momenta of the bosons are listed in \texttt{TransversePolarizationVectors}. The splitting of fermion-fermion-boson couplings into left and right handed chirality projectors (default in \textsc{FeynArts}) can be undone with the option \texttt{UndoChiralSplittings}. For example, the amplitude for the tree level process $\gamma^\ast \, u \to u \, g$ is obtained via

\begin{lstlisting}[language=Mathematica,mathescape]
In[1]:= $\$$LoadFeynArts = True;
        $\$$FeynCalcStartupMessages = False;
        << FeynCalc`;
        $\$$FAVerbose = 0;

In[2]:= diags = InsertFields[ CreateTopologies[0, 2 -> 2], {F[3, {1}],
         V[1]} -> {V[5], F[3, {1}]},  InsertionLevel -> {Classes},
         Model -> "SMQCD"];

In[3]:= FCFAConvert[CreateFeynAmp[diags], IncomingMomenta -> {p1, kp},
        OutgoingMomenta -> {kg, p2}, UndoChiralSplittings -> True,
        TransversePolarizationVectors -> {kg}, DropSumOver -> True,
         List -> False] // Contract

Out[3]:= $-\frac{2 \text{EL} g_s T_{\text{Col4}\text{Col1}}^{\text{Glu3}} \left(\varphi (\overline{\text{p2}},\text{MU})\right).\left(\bar{\gamma }\cdot \bar{\varepsilon }^*(\text{kg})\right).\left(\bar{\gamma }\cdot \left(\overline{\text{kg}}+\overline{\text{p2}}\right)+\text{MU}\right).\left(\bar{\gamma }\cdot \bar{\varepsilon }(\text{kp})\right).\left(\varphi (\overline{\text{p1}},\text{MU})\right)}{3 \left((-\text{kg}-\text{p2})^2-\text{MU}^2\right)}$
         $-\frac{2 \text{EL} g_s T_{\text{Col4}\text{Col1}}^{\text{Glu3}} \left(\varphi (\overline{\text{p2}},\text{MU})\right).\left(\bar{\gamma }\cdot \bar{\varepsilon }(\text{kp})\right).\left(\bar{\gamma }\cdot \left(\overline{\text{p2}}-\overline{\text{kp}}\right)+\text{MU}\right).\left(\bar{\gamma }\cdot \bar{\varepsilon }^*(\text{kg})\right).\left(\varphi (\overline{\text{p1}},\text{MU})\right)}{3 \left((\text{kp}-\text{p2})^2-\text{MU}^2\right)}$

\end{lstlisting}

\subsection{Finer-grained expansions}
To expand scalar products of Lorentz vectors \textsc{FeynCalc} provides the function \texttt{ExpandScalarProduct}. The standard behavior of this command is to expand every scalar product in the expression.

\begin{lstlisting}[language=Mathematica,mathescape]
In[1]:=  exp = SPD[q1, p1 + p2] SPD[q2, p3 + p4] SPD[p5 + p6, p7 + p8]

Out[1]:=  $((\text{p1}+\text{p2})\cdot \text{q1}) ((\text{p3}+\text{p4})\cdot \text{q2}) ((\text{p5}+\text{p6})\cdot (\text{p7}+\text{p8}))$

In[2]:=  ExpandScalarProduct[exp]

Out[2]:= $(\text{p1}\cdot \text{q1}+\text{p2}\cdot \text{q1}) (\text{p3}\cdot \text{q2}+\text{p4}\cdot \text{q2}) (\text{p5}\cdot \text{p7}+\text{p5}\cdot \text{p8}+\text{p6}\cdot \text{p7}+\text{p6}\cdot \text{p8})$

\end{lstlisting}
which might lead to an unnecessary increase of terms, if the user wants to expand only some particular scalar products. \textsc{FeynCalc} 9.0 improves \texttt{ExpandScalarProduct} by introducing the option \texttt{Momentum} which allows to specify a list of momenta that need to be contained in a scalar product that will be expanded. All the other scalar products will remain untouched.

\begin{lstlisting}[language=Mathematica,mathescape]
In[1]:=  exp = SPD[q1, p1 + p2] SPD[q2, p3 + p4] SPD[p5 + p6, p7 + p8]

Out[1]:=  $((\text{p1}+\text{p2})\cdot \text{q1}) ((\text{p3}+\text{p4})\cdot \text{q2}) ((\text{p5}+\text{p6})\cdot (\text{p7}+\text{p8}))$

In[2]:=  ExpandScalarProduct[exp, Momentum -> {q1}]

Out[2]:= $(\text{p1}\cdot \text{q1}+\text{p2}\cdot \text{q1}) ((\text{p3}+\text{p4})\cdot \text{q2}) ((\text{p5}+\text{p6})\cdot (\text{p7}+\text{p8}))$

In[3]:=  ExpandScalarProduct[exp, Momentum -> {q2}]

Out[2]:= $(\text{p3}\cdot \text{q2}+\text{p4}\cdot \text{q2}) ((\text{p1}+\text{p2})\cdot \text{q1}) ((\text{p5}+\text{p6})\cdot (\text{p7}+\text{p8}))$


\end{lstlisting}

The same option is now present also in \texttt{DiracGammaExpand} that is used to expand Lorentz vectors contracted with Dirac matrices

\begin{lstlisting}[language=Mathematica,mathescape]
In[1]:=  exp = GSD[q1 + p1 + p2].GSD[q2 + p3 + p4].GSD[p5 + p6 + p7 + p8]

Out[1]:=  $(\gamma \cdot (\text{p1}+\text{p2}+\text{q1})).(\gamma \cdot (\text{p3}+\text{p4}+\text{q2})).(\gamma \cdot (\text{p5}+\text{p6}+\text{p7}+\text{p8}))$

In[2]:=  DiracGammaExpand[exp]

Out[2]:= $(\gamma \cdot \text{p1}+\gamma \cdot \text{p2}+\gamma \cdot \text{q1}).(\gamma \cdot \text{p3}+\gamma \cdot \text{p4}+\gamma \cdot \text{q2}).(\gamma \cdot \text{p5}+\gamma \cdot \text{p6}+\gamma \cdot \text{p7}+\gamma \cdot \text{p8})$

In[3]:=  DiracGammaExpand[exp, Momentum -> {q1}]

Out[3]:=  $(\gamma \cdot \text{p1}+\gamma \cdot \text{p2}+\gamma \cdot \text{q1}).(\gamma \cdot (\text{p3}+\text{p4}+\text{q2})).(\gamma \cdot (\text{p5}+\text{p6}+\text{p7}+\text{p8}))$

In[4]:=  DiracGammaExpand[exp, Momentum -> {q2}]

Out[4]:=  $(\gamma \cdot (\text{p1}+\text{p2}+\text{q1})).(\gamma \cdot \text{p3}+\gamma \cdot \text{p4}+\gamma \cdot \text{q2}).(\gamma \cdot (\text{p5}+\text{p6}+\text{p7}+\text{p8}))$

\end{lstlisting}

\subsection{\texorpdfstring{$SU(N)$}{SUN} generators with explicit fundamental indices}

\textsc{FeynCalc} denotes $SU(N)$ generators in the fundamental representation as \texttt{SUNT[a]} where \texttt{a} stands for the adjoint index. The fundamental indices are suppressed, so that a chain of \texttt{SUNT}-matrices is understood to have only two free fundamental indices, e.g. \texttt{SUNT[a,b,c]} stands for $T^a_{ij} T^b_{jk} T^c_{kl}$ and it is not possible to express, say $T^a_{ij} T^b_{kl}$ with \texttt{SUNT} objects only.

Due to this limitation, evaluation of Feynman amplitudes with more than two free fundamental color indices (e.g $q \bar{ q} \to  q \bar{q}$ scattering in QCD) was very inconvenient and usually required additional \textsc{Mathematica} code to obtain the correct result. For this reason \textsc{FeynCalc} 9.0 introduces a new object \texttt{SUNTF[\{a\},i,j]} that stands for $T^a_{ij}$, an $SU(N)$ generator in the fundamental representation with explicit fundamental indices \texttt{i} and \texttt{j} and the adjoint index \texttt{a}. Hence expressions like $T^a_{ij} T^b_{kl}$ or  $T^a_{ij} T^b_{jk} T^c_{lm}$ can be now conveniently expressed with \texttt{SUNTF[\{a\},i,j]*SUNTF[\{b\},k,l]} and \texttt{SUNTF[\{a,b\},i,k]*SUNTF[\{c\},l,m]} respectively.
The new \texttt{SUNTF} objects are fully compatible with \texttt{SUNSimplify}, the standard routine for simplifying $SU(N)$ algebra.

\begin{lstlisting}[language=Mathematica,mathescape]
In[1]:=  exp1 = SUNTF[{a}, i, j] SUNTF[{b}, j, k] SUNTF[{c}, k, l]

Out[1]:=  $T_{ij}^a T_{jk}^b T_{kl}^c$

In[2]:=  SUNSimplify[exp1]

Out[2]:= $\left(T^aT^bT^c\right){}_{il}$

In[3]:=  exp2 = exp1 SUNFDelta[i, l]

Out[3]:= $\delta _{il} T_{ij}^a T_{jk}^b T_{kl}^c$

In[4]:=  SUNSimplify[exp2]

Out[4]:=  $\text{tr}(T^c.T^a.T^b)$

\end{lstlisting}

\section{Using FeynCalc with non-relativistic EFTs}

Up to now  we silently assumed that all the amplitudes and expressions that we want to evaluate stem from a theory that is manifestly Lorentz covariant. This nice property of relativistic QFTs is often taken for granted, but one surely should not forget about EFTs that are used to describe non-relativistic  systems, where the corresponding Lagrangians often do not exhibit manifest Lorentz covariance.

To our knowledge, there are no public tools for doing algebraic calculations in non-relativistic EFTs, where one has to explicitly distinguish between temporal and spatial components of Lorentz tensors. Naively, one might think that to do a calculation in such a theory using computer, one would need to write a large amount of additional code almost from scratch. However, with such a versatile tool like \textsc{FeynCalc}, this estimate turns out
to be too pessimistic. In the following we want to give a simple example of using \textsc{FeynCalc} in a non-relativistic calculation, where only a comparably small amount of additional \textsc{Mathematica} code is needed.

In Sec. \ref{sec:comparison} we have already mentioned NRQCD \cite{Bodwin1995}, which is an EFT of QCD that was developed to exploit the separation of scales
\begin{equation}
m v^2 \ll mv \ll m
\end{equation}
in a heavy quarkonium. Here, $m$ denotes the heavy quark mass and $v$ stands for the relative velocity of heavy quarks in the quarkonium. The scales $m$, $mv$ and $mv^2$ are usually called hard, soft and ultrasoft respectively.

NRQCD is obtained from QCD by integrating out all degrees of freedom above the soft scale. The hard contributions are of course not simply thrown away. Their effects are incorporated in the matching coefficients $\omega_n$ that multiply operators $\mathcal{O}_n$ of the NRQCD Lagrangian, which can be schematically written as
\begin{equation}
\mathcal{L}_{\textrm{NRQCD}} = \sum_n \frac{\omega_n}{m^n} \mathcal{O}_n.
\end{equation}
Since for charm and bottom quarks we have
\begin{equation}
m \gg \Lambda_{\textrm{QCD}},
\end{equation}
with $\Lambda_{\textrm{QCD}}$ being the QCD scale at which the perturbation theory breaks down, the matching can be always done perturbatively.

The matching coefficients are fixed by comparing suitable quantities in perturbative QCD and in perturbative NRQCD at finite order in the expansion in $v$. The NRQCD Lagrangian itself contains an infinite number of operators of arbitrary high dimensions that are compatible with the symmetries of QCD. Using the power counting rules of the theory, one can estimate the relative importance of the operators for each process of interest. For this reason, usually only a small number of NRQCD operators needs to be considered in a practical calculation.

In the following we want to use \textsc{FeynCalc} to perform the matching between QCD and NRQCD in order to extract the matching coefficients (at leading order in $\alpha_s$) that enter the decay rate of $\chi_{c_{0,2}} \to \gamma \gamma$  at leading order in $v$. Notice that the decay $\chi_{c_1} \to \gamma \gamma$ does not occur, because it is forbidden by the Landau-Yang theorem.

These matching coefficients have been calculated in the framework of NRQCD multiple times \cite{Bodwin1995,Petrelli1998,Ma2002,Mereghetti2006,Sang2015}, with many of these calculations carried out in a fully covariant way using the covariant projector technique \cite{Bodwin2002}. Nevertheless, for pedagogical reasons we want to stick to the explicit non-covariant matching in the spirit of \cite{Bodwin1995} and \cite{Mereghetti2006}. We also would like to remark that the projector technique has not yet been generalized for higher quarkonium Fock states, that include not only two heavy quarks $\ket{Q \bar{Q}}$ but also gluons (e.g. $\ket{Q \bar{Q} g}$ or $\ket{Q \bar{Q} gg}$). For this reason, the presented approach might still be useful in calculations, where such higher order contributions have to be considered. We also want to stress that codes which offer out of the box support for doing NRQCD calculations already exist (e.g. \textsc{FDC} \cite{Wang2004} package), so the current example merely shows a quick naive implementation not optimized for performance or flexibility.

The factorization formulas for the decay rates \cite{Bodwin1995} are given by

\begin{align}
\Gamma (\chi_{c_0} \to \gamma \gamma) &= \frac{2\textrm{Im} f_{em} (^3 P_0)}{ 3m^4} \braket{\chi_{c_0}| \chi^\dagger ( - \tfrac{i}{2} \overleftrightarrow{\bfD} \cdot \bfSigma )  \psi |0} \braket{0| \psi^\dagger ( - \tfrac{i}{2} \overleftrightarrow{\bfD} \cdot \bfSigma )  \chi |\chi_{c_0}} \\
\Gamma (\chi_{c_2} \to \gamma \gamma) &= \frac{2\textrm{Im} f_{em} (^3 P_2)}{m^4} \braket{\chi_{c_2}| \chi^\dagger ( - \tfrac{i}{2} \overleftrightarrow{\bfD}^{(i} \bfSigma^{j)} )  \psi |0} \braket{0| \psi^\dagger ( - \tfrac{i}{2} \overleftrightarrow{\bfD}^{(i}  \bfSigma^{j)} )  \chi |\chi_{c_2}}.
\end{align}
Here, Pauli spinor field $\psi$ ($\chi$) annihilates (creates) a heavy quark (antiquark), $\bfSigma$ is the Pauli vector
and the covariant derivative is defined as 
\begin{align}
D^\mu = \partial^\mu + i g A^\mu \equiv (D^0,-\bfD),
\end{align}
so that
\begin{align}
i D^0 & = i \partial^0 - g A^0, \\
i \bfD & = i \bfDel + g \bfA,
\end{align}
where $A^\mu$ is the gluon field and $g$ stands for the QCD coupling constant. Furthermore,
\begin{align}
\psi^\dagger \stackrel{\leftrightarrow}{\bfD} \chi & \equiv \psi^\dagger (\bfD \chi) - (\bfD \psi)^\dagger\chi, \\
\overleftrightarrow{\bfD}^{(i} \bfSigma^{j)} & \equiv \frac{1}{2} \left ( \overleftrightarrow{\bfD}^i \bfSigma^j + \overleftrightarrow{\bfD}^j \bfSigma^i \right ) - \frac{1}{3} \delta^{ij} \overleftrightarrow{\bfD} \cdot \bfSigma.
\end{align}
The NRQCD long distance  matrix elements (LDME)
\begin{equation}
\braket{\chi_{c_0}| \chi^\dagger ( - \tfrac{i}{2} \overleftrightarrow{\bfD} \cdot \bfSigma )  \psi |0} \braket{0| \psi^\dagger ( - \tfrac{i}{2} \overleftrightarrow{\bfD} \cdot \bfSigma )  \chi |\chi_{c_0}}
\end{equation}
and
\begin{equation}
 \braket{\chi_{c_2}| \chi^\dagger ( - \tfrac{i}{2} \overleftrightarrow{\bfD}^{(i} \bfSigma^{j)} )  \psi |0} \braket{0| \psi^\dagger ( - \tfrac{i}{2} \overleftrightarrow{\bfD^{(i} }  \bfSigma^{j)} )  \chi |\chi_{c_2}}
\end{equation}
are non-perturbative. They can be determined from fitting to the experimental data or computed on the lattice. On the other hand, the matching coefficients $f_{em} (^3 P_0)$ and $f_{em} (^3 P_2)$ can be calculated in perturbation theory from the matching condition \cite{Bodwin1995}
\begin{align}
& 2 \, \textrm{Im} \, A \, (Q \bar{Q} \to Q \bar{Q})  \biggl |_{\textrm{pert. QCD}} \nonumber \\
&= \frac{2\textrm{Im} f_{em} (^3 P_0)}{3m^4} \braket{Q \bar{Q}| \chi^\dagger ( - \tfrac{i}{2} \overleftrightarrow{\bfD} \cdot \bfSigma )  \psi |0}  \braket{0| \psi^\dagger ( - \tfrac{i}{2} \overleftrightarrow{\bfD} \cdot \bfSigma )  \chi | Q \bar{Q}} |_{\textrm{pert. NRQCD}}  \nonumber \\
& +  \frac{ 2\textrm{Im} f_{em} (^3 P_2)}{m^4} \braket{Q \bar{Q}| \chi^\dagger ( - \tfrac{i}{2} \overleftrightarrow{\bfD}^{(i} \bfSigma^{j)} )  \psi |0} \braket{0| \psi^\dagger ( - \tfrac{i}{2} \overleftrightarrow{\bfD}^{(i}  \bfSigma^{j)} )  \chi |Q \bar{Q}} |_{\textrm{pert. NRQCD}} 
\label{eq:matchingCondition},
\end{align}
where on the right hand side we have displayed only spin triplet terms that contribute at leading order in $v$. The left hand side of Eq. \ref{eq:matchingCondition} denotes twice the imaginary part of the perturbative QCD amplitude $Q \bar{Q} \to  Q \bar{Q}$ with 2 photons in the intermediate state. It is understood that this amplitude also has to be expanded to second order in $v$.

We start the matching calculation by considering the on-shell amplitude for the perturbative process $Q (p_1) \bar{Q}(p_2) \to \gamma (k_1) \gamma (k_2)$ in QCD.  The kinematics of this process reads
\begin{align}
 p_1 + p_2 & = k_1 + k_2, \\
 p_1^2 & = p_2^2 = m^2, \\
 k_1^2 & = k_2^2 = 0,
\end{align}
with $p_{i} = (\sqrt{m^2 + \bfp_{i}^2}, \bfp_{i}) \equiv (E_{i}, \bfp_{i})$ and $k_{i} = (|\bfk_{i}|, \bfk_{i})$. 
Obviously, it is most convenient to work in the quarkonium rest frame, where
\begin{align}
\bfp_1 &= - \bfp_2 \equiv \bfq, \\
E_1 &= E_2  \equiv E_q = \sqrt{m + \bfq^2}, \\
\bfk_1 &= -\bfk_2.
\end{align}
For convenience, the photon polarization vectors can be chosen to be purely spatial, satisfying
\begin{align}
\epsilon (k_1)^0 &= \epsilon (k_2)^0  = 0, \\
\bfEps (k_{1/2}) \cdot \bfk_{1/2} & = \bfEps (k_{1/2}) \cdot \bfk_{2/1} = 0.
\end{align}
We need to expand the QCD amplitude in $v$, i.e. in $|\bfq|/m$ up to second order which involves rewriting
Dirac spinors for $Q$ and $\bar{Q}$ in terms of the Pauli spinors. For the latter let us recall that in 4-dimensions we can decompose any chain of Dirac matrices into scalar, pseudoscalar, vector, axial vector and tensor (SPVAT) components. This decomposition stems from the fact that the 4 dimensional matrices $I$, $\gamma^5$, $\gamma^\mu$, $\gamma^5 \gamma^\mu$ and $\sigma^{\mu \nu} = \frac{i}{2} [\gamma^\mu,\gamma^\nu]$ form a basis, such that any $4 \times 4$ matrix $M$ can be written as 
\begin{equation}
M = c_1 I + c_2 \gamma^5 + c_{3\mu} \gamma^\mu + c_{4\mu} \gamma^5 \gamma^\mu + c_{5 \mu \nu} \sigma^{\mu \nu}.
\end{equation}

Therefore, there are only 5 unique spinor structures involving heavy quarks that we can encounter in any tree level amplitude.
In fact, the only components that appear in this calculation are vector and axial vector, so that we do not need to consider the other three. Using the explicit form of the Dirac spinors with the non-relativistic normalization, 

\begin{align}
u(\bfq) &= \sqrt{\frac{E_{q} + m}{2 E_q}}\begin{pmatrix}
  \xi  \\  \frac{\bfq \cdot \bfSigma }{E_{q} + m}\xi
\end{pmatrix},  \\
v(-\bfq) &= \sqrt{\frac{E_{q} + m}{2 E_q}}\begin{pmatrix}
 - \frac{ \bfq \cdot \bfSigma }{E_{q} + m} \eta  \\  \eta
\end{pmatrix},
\end{align}
with $\xi$ and $\eta$ being 2-component spinors, we obtain 

\begin{align}
\bar{v}(-\bfq) \gamma^0 u(\bfq) &= 0, \label{eq:chainsFirst} \\ 
\bar{v}(-\bfq) \gamma^i v(\bfq) &=  \eta^\dagger\bfSigma^i \xi - \frac{\bfq^i}{2 m^2}  \eta^\dagger \bfq \cdot \bfSigma \xi + \mathcal{O}((|\bfq|/m)^3), \\
\bar{v}(-\bfq) \gamma^0 \gamma^5 u(\bfq) &=   \eta^\dagger \xi \left (1 -\frac{\bfq^2}{2 m^2} \right )  + \mathcal{O}((|\bfq|/m)^3), \\ 
\bar{v}(-\bfq) \gamma^i \gamma^5 u(\bfq) &=
\frac{i}{m} \eta^\dagger (\bfq \times  \bfSigma)^i \xi  + \mathcal{O}((|\bfq|/m)^3) \label{eq:chainsLast}.
\end{align}

Let us first ignore all the complications related to the non-relativistic expansion and see how far we can get with the QCD amplitude without breaking the covariant notation.

At this order in $v$ and $\alpha_s$, there are only two tree level diagrams to consider that can be trivially generated with \textsc{FeynArts}.

\begin{lstlisting}[language=Mathematica,mathescape]
In[1]:= $\$$LoadFeynArts = True;
        $\$$FeynCalcStartupMessages = False;
        << FeynCalc`;
        $\$$FAVerbose = 0;

In[2]:= diags = InsertFields[CreateTopologies[0, 2 -> 2],
	   {F[3, {2, a}], -F[3, {2, b}]} -> {V[1], V[1]},
	   InsertionLevel -> {Classes}, Model -> "SMQCD"];
\end{lstlisting}
Then the amplitudes are converted into \textsc{FeynCalc} notation and simplified using standard \textsc{FeynCalc} functions.
\begin{lstlisting}[language=Mathematica,mathescape]
In[3]:= amps = (9/4 EQ^2*FCFAConvert[
		CreateFeynAmp[diags, Truncated -> False, PreFactor -> -1],
		IncomingMomenta -> {p1, p2}, OutgoingMomenta -> {k1, k2},
		UndoChiralSplittings -> True, 
		TransversePolarizationVectors -> {k1, k2},
		ChangeDimension -> 4, List -> False]) // Contract // Factor

Out[3]:= $i \text{EL}^2 \text{EQ}^2 \delta _{ab} \frac{\left(\varphi (-\overline{\text{p2}},\text{MC})\right).\left(\bar{\gamma }\cdot \bar{\varepsilon }^*(\text{k1})\right).\left(\bar{\gamma }\cdot \left(\overline{\text{k1}}-\overline{\text{p2}}\right)+\text{MC}\right).\left(\bar{\gamma }\cdot \bar{\varepsilon }^*(\text{k2})\right).\left(\varphi (\overline{\text{p1}},\text{MC})\right)}{\left(\overline{\text{p2}}-\overline{\text{k1}}\right)^2-\text{MC}^2}$
				$+i \text{EL}^2 \text{EQ}^2 \delta _{ab} \frac{\left(\varphi (-\overline{\text{p2}},\text{MC})\right).\left(\bar{\gamma }\cdot \bar{\varepsilon }^*(\text{k2})\right).\left(\bar{\gamma }\cdot \left(\overline{\text{k2}}-\overline{\text{p2}}\right)+\text{MC}\right).\left(\bar{\gamma }\cdot \bar{\varepsilon }^*(\text{k1})\right).\left(\varphi (\overline{\text{p1}},\text{MC})\right)}{\left(\overline{\text{p2}}-\overline{\text{k2}}\right)^2-\text{MC}^2}$

\end{lstlisting}
The next step is to put the external particles on-shell

\begin{lstlisting}[language=Mathematica,mathescape]
In[4]:=  FCClearScalarProducts[];
					ScalarProduct[k1, k1] = 0;
					ScalarProduct[k2, k2] = 0;
					ScalarProduct[p1, p1] = MC^2;
					ScalarProduct[p2, p2] = MC^2;
\end{lstlisting}
and perform the SPVAT decomposition of the spinor chains,
\begin{lstlisting}[language=Mathematica,mathescape]
In[5]:=  repRuleHideChains = {
   FCI[Spinor[-p2, MC].GA[x_].GA[5].Spinor[p1, MC]] :> FCI[FV[A, x]],
   FCI[Spinor[-p2, MC].GA[x_].Spinor[p1, MC]] :> FCI[FV[V, x]],
   FCI[Spinor[-p2, MC].GS[x_].GA[5].Spinor[p1, MC]] :> FCI[SP[A, x]],
   FCI[Spinor[-p2, MC].GS[x_].Spinor[p1, MC]] :> FCI[SP[V, x]]
   };
   
In[6]:=  amps2 = amps // DiracSimplify // DiracReduce // FCI // 
		     ReplaceAll[#, repRuleHideChains] & //    
		     PropagatorDenominatorExplicit[#, Dimension -> 4] & // 
		     Contract // ReplaceAll[#, Pair[Momentum[k1 | k2], 
		     Momentum[Polarization[k1 | k2, ___]]] -> 0] &   
		     
Out[6]:= $\frac{\text{EL}^2 \text{EQ}^2 \delta _{ab} \epsilon ^{\overline{A}\overline{\text{k1}}\bar{\varepsilon }^*(\text{k1})\bar{\varepsilon }^*(\text{k2})}}{2 \left(\overline{\text{k1}}\cdot \overline{\text{p2}}\right)}-\frac{\text{EL}^2 \text{EQ}^2 \delta _{ab} \epsilon ^{\overline{A}\overline{\text{k2}}\bar{\varepsilon }^*(\text{k1})\bar{\varepsilon }^*(\text{k2})}}{2 \left(\overline{\text{k2}}\cdot \overline{\text{p2}}\right)}+\frac{i \text{EL}^2 \text{EQ}^2 \delta _{ab} \left(\overline{V}\cdot \bar{\varepsilon }^*(\text{k1})\right) \left(\overline{\text{p2}}\cdot \bar{\varepsilon }^*(\text{k2})\right)}{\overline{\text{k2}}\cdot \overline{\text{p2}}}$
+$\frac{i \text{EL}^2 \text{EQ}^2 \delta _{ab} \left(\overline{\text{p2}}\cdot \bar{\varepsilon }^*(\text{k1})\right) \left(\overline{V}\cdot \bar{\varepsilon }^*(\text{k2})\right)}{\overline{\text{k1}}\cdot \overline{\text{p2}}}+\frac{i \text{EL}^2 \text{EQ}^2 \delta _{ab} \left(\overline{\text{k1}}\cdot \overline{V}\right) \left(\bar{\varepsilon }^*(\text{k1})\cdot \bar{\varepsilon }^*(\text{k2})\right)}{2 \left(\overline{\text{k1}}\cdot \overline{\text{p2}}\right)}+\frac{i \text{EL}^2 \text{EQ}^2 \delta _{ab} \left(\overline{\text{k2}}\cdot \overline{V}\right) \left(\bar{\varepsilon }^*(\text{k1})\cdot \bar{\varepsilon }^*(\text{k2})\right)}{2 \left(\overline{\text{k2}}\cdot \overline{\text{p2}}\right)}$

\end{lstlisting}
where for convenience we chose to abbreviate vector and axial vector chains as 
\begin{align}
V^\mu &  \equiv \bar{v}(\bfp_2) \gamma^\mu u(\bfp_1), \\ 
A^\mu &  \equiv \bar{v}(\bfp_2) \gamma^\mu \gamma^5 u(\bfp_1).
\end{align}
If we are to expand the resulting expression in $|\bfq|/m$, we must make the $\bfq$-dependence explicit in all parts of the amplitude. Since different components of the 4-vectors and spinor chains that appear in the computation depend on $|\bfq|$ in a different way, it now becomes necessary to break the covariant notation.  However, by doing so in a naive way, e.g. by writing something like
\begin{align}
 V \cdot k_1 & = V^0 |\bfk| - \bfV \cdot \bfk, \\
\epsilon^{\mu \nu \rho \sigma} k_{1 \mu} A_{\nu} \varepsilon_\rho^\ast(k_1) \varepsilon_\sigma^\ast(k_2)  &= 
\epsilon^{\mu 0 \rho \sigma} k_{1 \mu} A^0 \varepsilon_\rho^\ast(k_1) \varepsilon_\sigma^\ast(k_2) - \nonumber \\
& \epsilon^{\mu i \rho \sigma} k_{1 \mu} A^i \varepsilon_\rho^\ast(k_1) \varepsilon_\sigma^\ast(k_2)  
\end{align}
we introduce new objects that carry Cartesian indices and thus cannot be handled by the built-in routines for working with Lorentz tensors (e.g. \texttt{Con\-tract}, \texttt{Scalar\-Product}, \texttt{ExpandScalar\-Product} etc.). Fortunately, it is possible to completely avoid introducing any Cartesian tensors  or tensors with mixed Lorentz and Cartesian indices by exploiting \textsc{FeynCalc}'s built-in \texttt{TensorFunction} in a clever way. 
 
This approach is based on \cite{Braaten1996a}, although we do not consider a boosted $Q \bar{Q}$-system and assume that the  quarkonium is at rest. To see how this works, let us first define a symmetric tensor $E^{\mu \nu}$ with
\begin{align}
E^{\mu \nu} =
\begin{cases}
 0  \mbox{ for } \mu = 0 \mbox{ or } \nu = 0, \\
 \delta^{ij}   \mbox{ for } \mu \neq 0 \mbox{ and } \nu \neq 0 .
\end{cases}
\end{align} 
With $E^{\mu \nu}$ we can write any Cartesian scalar product  $x^i y^i$ as 
\begin{equation}
x^i y^i = x^i y^j \delta^{ij} =   E^{\mu \nu} x_{\mu} y_{\nu} \equiv E(x,y),
\end{equation}
where $x^0$ and $y^0$ can be anything, since they drop out by construction. If $x$ is a pure Cartesian vector, then we can choose $x^\mu = (0,x^i)$.  Suppose that we want to expand $x^i y^i$ in $|\bfx|$ or $|\bfy|$. Then we can write
\begin{equation}
E (x,y) = |\bfx| |\bfy| E(\hat{x}, \hat{y}),
\end{equation}
with $x^i = |\bfx| \hat{x}^i$ and $y^i = |\bfy|  \hat{y}^i$, where $E(\hat{x}, \hat{y})$ does not depend on $|\bfx|$ and $|\bfy|$. Therefore, a Minkowski scalar product $x \cdot y$ can be rewritten as
\begin{equation}
x^\mu y_\mu = x^0 y^0  + |\bfx| |\bfy| E (\hat{x},\hat{y})
\end{equation}
and if $x$ and $y$ are external 4-momenta, then we have
\begin{equation}
x^\mu y_\mu = \sqrt{m_x^2 + |\bfx|^2} \sqrt{m_y^2 + |\bfy|^2}  + |\bfx| |\bfy| E (\hat{x},\hat{y}),
\end{equation}
so that the expansion in the scalar variables $|\bfx|$ or $|\bfy|$ can be carried out without making any reference to 3-vectors.  Some useful relations for dealing with $E$-tensors are
\begin{align}
E^{\mu \nu} g_{\mu \nu} &= -3, \\
E^{\mu \nu} E^{\rho \sigma} g_{\mu \rho}  &= - E^{\nu \sigma}
\end{align}
In a similar manner we can also rewrite terms that involve 3-dimensional epsilon tensors by introducing
\begin{equation}
C_{\mu \nu \rho} =
\begin{cases}
  0  \mbox{ for } \mu = 0 \mbox{ or } \nu = 0 \mbox{ or }  \rho = 0, \\
\varepsilon_{ijk}  \mbox{ for } \mu \neq 0 \mbox{ and } \nu \neq 0 \mbox{ and }  \rho \neq 0,
\end{cases}
\end{equation}
such that 
\begin{equation}
\varepsilon^{ijk} x^i y^j z^k =  - \varepsilon_{ijk} x^i y^j z^k = - C_{\mu \nu \rho} x^\mu y^\nu z^\rho \equiv - C(x,y,z).
\end{equation}
Then, it is easy to see that
\begin{align}
\varepsilon^{\sigma \mu \nu \rho} a_\sigma x_\mu y_\nu z_\rho & = \varepsilon^{ijk} (a^0 x_i y_j z_k  - x^0 a_i y_j z_k + y^0 a_i x_j z_k  - z^0 a_i x_j y_k ) \nonumber \\
& = a^0 C(x,y,z) - x^0 C(a,y,z) + y^0 C(a,x,z) - z^0 C(a,x,y) \label{eq:nrexpand},
\end{align}
from where we again can easily expand in the 3-momenta of $a$, $x$, $y$ or $z$, since
\begin{equation}
C(x,y,z) = |\bfx| |\bfy| |\bfz| C(\hat{x},\hat{y},\hat{z}),
\end{equation}
with $C(\hat{x},\hat{y},\hat{z})$ being independent of $|\bfx|$, $|\bfy|$ and $|\bfz|$. 
The product of two $C$ tensors can be expressed through
\begin{align}
C^{\mu \nu \rho} C^{\alpha \beta \gamma } = \begin{vmatrix}
  E^{\mu \alpha} & E^{\mu \beta} & E^{\mu \gamma}   \\   E^{\nu \alpha} & E^{\nu \beta} & E^{\nu \gamma} \\  E^{\rho \alpha} & E^{\rho \beta} & E^{\rho \gamma} \end{vmatrix}.
\end{align}
The basic properties of $E^{\mu \nu}$ (denoted as \texttt{NRPair}) and $C^{\mu \nu \rho}$ (denoted as \texttt{NREps}) can be implemented in \textsc{FeynCalc} with a minimal amount of extra code.

\begin{lstlisting}[language=Mathematica,mathescape]
In[7]:=  SetAttributes[NRPairContract, Orderless];
					TensorFunction[NREps, x, y, z];
					TensorFunction[{NRPair, "S"}, x, y];
					NREps[a___, x_, b___, x_, c___] := 0;
					NRPairContract /:	
					 NRPairContract[LorentzIndex[x_], LorentzIndex[x_]] := -3;
					NRPairContract /:
					 NRPairContract[LorentzIndex[x_], y_] *
					 NRPairContract[LorentzIndex[x_], z_] := - NRPairContract[y, z];
					NRPairContract /: 
					 NRPairContract[LorentzIndex[x_], y_] *
					 NREpsContract[a___, LorentzIndex[x_], b___] := - NREpsContract[a, y, b];
					NRPairContract /:
 					 NRPairContract[LorentzIndex[x_], y_]^2 := - NRPairContract[y, y];
					NREpsContract /:
					 NREpsContract[x_, y_, z_]^2 := - 6;
					 NREpsContract /:  
				  NREpsContract[mu_, nu_, rho_] NREpsContract[al_, be_, ga_] := 
				     (Det[{{np[ mu, al], np[ mu, be], np[ mu, ga]},
				      {np[ nu, al], np[ nu, be], np[nu, ga]},
				      {np[ rho, al], np[ rho, be], np[ rho, ga]}}] /. 
					    np -> NRPairContract);
\end{lstlisting}
Contractions of $E^{\mu \nu}$ and $C^{\mu \nu \rho}$ with each other or with the metric tensor are simplified by  \texttt{NRContract}, while \texttt{NRExpand} implements Eq. \ref{eq:nrexpand}.
\begin{lstlisting}[language=Mathematica,mathescape]
In[8]:=  NRContract[expr_] := 
				  FixedPoint[(Expand2[Contract[#], {NRPair, NREps}] //. {NRPair -> 
				  NRPairContract, NREps -> NREpsContract}) &, expr] /. 
				  {NRPairContract -> NRPair, NREpsContract -> NREps};
				  
In[9]:=  NRExpand[expr_] := 
		  FixedPoint[ReplaceRepeated[Expand2[#, {Eps, NREps}], 
		  {Eps[a_Momentum, x_Momentum, y_Momentum, z_Momentum] :> 
       	NREn[a] NREps[x, y, z] - NREn[x ] NREps[a, y, z] + 
        NREn[y] NREps[a, x, z] - NREn[z] NREps[a, x, y]}] &, expr];				  
				  
\end{lstlisting}
Here we use \texttt{NREn} to denote the temporal components of 4-momenta. The expansions of spinor chains given in Eqs. \ref{eq:chainsFirst} - \ref{eq:chainsLast} are now straight-forward to translate into \textsc{FeynCalc} notation.

\begin{lstlisting}[language=Mathematica,mathescape]
In[10]:=  repRuleExpandedChains = {
			   Pair[v : Momentum[V], x_] :> -NRPair[v, x],
			   Pair[a : Momentum[A], x_] :> NREn[a] NREn[x] - NRPair[a, x],
			   NREn[Momentum[A]] -> 1 - (qvec^2) /(2 MC^2),
			   NREn[Momentum[V]] -> 0,
			   NRPair[x_, Momentum[V]] :>  -((qvec^2 NRPair[Momentum[qhat], 
	           Momentum[{S, I}]] NRPair[Momentum[qhat], x])/(2 MC^2)) +
		     NRPair[Momentum[{S, I}], x],
			   NRPair[x_, Momentum[A]] :> 
	   		-((I qvec NREps[Momentum[qhat], Momentum[{S, I}], x])/MC),
			   NREps[x___, a : Momentum[A], y___] :> (li = 
		      LorentzIndex[Unique[]]; -NREps[x, li, y] NRPair[a, li]),
			   NREps[x___, v : Momentum[V], y___] :> 
	   		(li = LorentzIndex[Unique[]]; -NREps[x, li, y] NRPair[v, li])
   };
\end{lstlisting}
Simplifications that are specific to the kinematics of the process are also easy to define.

\begin{lstlisting}[language=Mathematica,mathescape]
In[11]:= NREn[Momentum[Polarization[k1 | k2, ___]]] = 0;
					NREn[Momentum[k1 | k2]] = kvec;
					NREn[Momentum[p1 | p2]] = Sqrt[MC^2 + qvec^2];
					NREn[Momentum[qhat | qhatp | {S, _}]] = 0;
					NREn[Momentum[k1hat]] = 1;
					NRPair[Momentum[p1], x_] = qvec NRPair[Momentum[qhat], x];
					NRPair[Momentum[p2], x_] = -qvec NRPair[Momentum[qhat], x];
					NRPair[Momentum[p1p], x_] = qvec NRPair[Momentum[qhatp], x];
					NRPair[Momentum[p2p], x_] = -qvec NRPair[Momentum[qhatp], x];
					NRPair[Momentum[k1 | k2 | k1hat | k2hat], 
						Momentum[Polarization[k2 | k1, ___]]] = 0;
					NRPair[Momentum[x_], Momentum[x_]] := 
				  1 /; MemberQ[{qhat, qhatp, p1hat, p2hat, k1hat, k2hat}, x];
					NRPair[Momentum[k1], x_] = kvec NRPair[Momentum[k1hat], x];
					NRPair[Momentum[k2], x_] = -kvec NRPair[Momentum[k1hat], x];
					kvec = Sqrt[MC^2 + qvec^2];

					repRuleExpansion = {
					   FCI@SP[x_, a : Polarization[z_, ___]] /; MemberQ[{k1, k2}, z] :>
					    -NRPair[Momentum[x], Momentum[a]],
					   FCI@SP[x_, (y : p1 | p2 | k1 | k2 | {S, I} | {S, -I} | k1hat | qhat)] :> 
					    NREn[Momentum[x]] NREn[Momentum[y]] -
					     NRPair[Momentum[x], Momentum[y]],
					   NREps[a___, Momentum[k1], z___] :> kvec NREps[a, Momentum[k1hat], z],
					   NREps[a___, Momentum[k2], z___] :> - kvec NREps[a, Momentum[k1hat], z],
					   NREps[a___, Momentum[p1], b___] :> qvec NREps[a, Momentum[qhat], b],
					   NREps[a___, Momentum[p2], b___] :> -qvec NREps[a, Momentum[qhat], b]
				   };
\end{lstlisting}
Finally, we can expand the amplitude up to second order in $|\bfq|/m$.
\begin{lstlisting}[language=Mathematica]
In[12]:=  amps3 = amps2 // NRExpand // 
					ReplaceRepeated[#, repRuleExpandedChains] & // NRContract // 
					ReplaceRepeated[#, repRuleExpansion] & // Series[#, {qvec, 0, 2}] & // 
					Normal // PowerExpand // NRContract;
\end{lstlisting}
To obtain $2 \, \textrm{Im} \, A  \, ( Q(p'_1) \bar{Q}(p'_2) \to Q(p_1) \bar{Q}(p_2))$ from our expanded amplitude, we need to multiply $A  \, ( Q(p'_1) \bar{Q}(p'_2) \to \gamma (k_1) \gamma (k_2) )$ by $A^\ast  \, ( Q(p_1) \bar{Q}(p_2) \to \gamma (k_1) \gamma (k_2) )$, sum over polarizations of the external photons and perform the phase space integration. For the latter we can use that
\begin{align}
\int d \Omega_{k_1} \, \hat{k}_1^{i_1} \ldots \hat{k}_1^{i_{2n+1}} &= 0, \\
\int d \Omega_{k_1} \, \hat{k}_1^{i_1} \ldots \hat{k}_1^{i_{2n}} &= \frac{4 \pi}{(n+2)!!} \left ( \delta^{i_1 i_2} \ldots \delta^{i_{2n-1} i_{2n}} + \textrm{permutations}  \right ).
\end{align}
To implement these relations we need an auxiliary function that uncontracts the indices of $\hat{k}_1$ 

\begin{lstlisting}[language=Mathematica]
In[13]:=  NRUncontract[expr_, l_List] :=  
  expr /. {Power[t_NRPair, n_] :> times @@ Table[t, {i, 1, n}],
      Power[t_NREps, n_] :> times @@ Table[t, {i, 1, n}]} //. {
     NRPair[y_, x_] /; ! FreeQ2[y, l] && Head[x] =!= LorentzIndex :> 
     (li = Unique[$AL]; -NRPair[y, LorentzIndex[li]] NRPair[x, LorentzIndex[li]]),
     NREps[w___, y_, x___] /; ! FreeQ2[y, l] :> (li = Unique[$AL]; 
     -NRPair[y, LorentzIndex[li]] NREps[w, LorentzIndex[li], x])
     } /. times -> Times;
\end{lstlisting}
and a replacement rule that handles the angular integration
\begin{lstlisting}[language=Mathematica]
In[14]:=  angularIntegration[hat_] := {
   qHead[NRPair[i_, Momentum[hat]]] /; FreeQ2[{i}, {hat, S}] :> 0,
   qHead[a_Times] :> qHead[(List @@ a) /. NRPair[Momentum[hat], b_] :> 
		   {hat, b /. LorentzIndex[c_, _] :> c}], 
   qHead[a_List] :> (Tdec[a, {}, List -> False, FCE -> False, 
       Dimension -> 3] //. {(h : LorentzIndex | Momentum)[x_, 3] :> 
        h[x], Pair -> NRPair})
   };
\end{lstlisting}
Then the left hand side of Eq. \ref{eq:matchingCondition} is given by
\begin{lstlisting}[language=Mathematica,mathescape]
In[15]:=  res = (1/(16 Pi)) (Collect[(amps3 /. qhat -> qhatp)*
                 ComplexConjugate[amps3] /. NRPair[x_, y_] :> 
                 -FCI@SP[x, y] + NREn[x] NREn[y], qvec] /. qvec^4 -> 0) // 
     		       		 DoPolarizationSums[#, k1, k2] & // 
		               DoPolarizationSums[#, k2, k1] & // 
		               ReplaceRepeated[#, repRuleExpansion] & // Cancel // 
		               NRUncontract[#, {k1hat}] & // 
		       	       FCLoopIsolate[#, {k1hat}, Head -> qHead] & // 
        			     ReplaceRepeated[#, angularIntegration[k1hat]] & // NRContract // 
      				     ReplaceAll[#, {EL^4 -> 16 Pi^2 AlphaFS ^2}] & // 
     				       SelectNotFree[#, S] &

Out[15]:=	$\frac{4 \pi  \alpha ^2 \text{EQ}^4 \text{qvec}^2 \delta _{a b}^2 \text{NRPair}\left(\overline{\text{qhat}},\overline{\{S,i\}}\right) \text{NRPair}\left(\overline{\text{qhatp}},\overline{\{S,-i\}}\right)}{5 \text{MC}^4}$+
					$\frac{22 \pi  \alpha ^2 \text{EQ}^4 \text{qvec}^2 \delta _{a b}^2 \text{NRPair}\left(\overline{\text{qhat}},\overline{\{S,-i\}}\right) \text{NRPair}\left(\overline{\text{qhatp}},\overline{\{S,i\}}\right)}{15 \text{MC}^4}$+
					$\frac{4 \pi  \alpha ^2 \text{EQ}^4 \text{qvec}^2 \delta _{a b}^2 \text{NRPair}\left(\overline{\text{qhat}},\overline{\text{qhatp}}\right) \text{NRPair}\left(\overline{\{S,-i\}},\overline{\{S,i\}}\right)}{5 \text{MC}^4}$						  
\end{lstlisting}
An explicit expression for the right hand side of Eq. \ref{eq:matchingCondition} can be obtained by using Fourier
decompositions of the Pauli spinor fields (c.f. \cite{Cho1996}), so that we end up with
\begin{align}
& \frac{4 \alpha^2 Q^4 \pi }{5 m^4} \bfq \cdot \bfq' \eta^\dagger \bfSigma \xi \, \xi^\dagger \bfSigma \eta + \frac{4 \alpha^2 Q^4 \pi}{5 m^4}  \eta^\dagger \bfq \cdot \bfSigma \xi \, \xi^\dagger \bfq' \cdot \bfSigma \eta  + \frac{22 \alpha^2 Q^4 \pi}{15 m^4}  \eta^\dagger \bfq' \cdot \bfSigma \xi \, \xi^\dagger \bfq \cdot \bfSigma \eta \nonumber \\
&= \frac{\textrm{Im} f_{em} (^3 P_2)}{m^4} \bfq \cdot \bfq' \, \eta^\dagger \bfSigma \xi \, \xi^\dagger \bfSigma \eta + \frac{\textrm{Im} f_{em} (^3 P_2)}{m^4}   \eta^\dagger \bfq \cdot \bfSigma \xi \, \xi^\dagger \bfq' \cdot \bfSigma \eta \nonumber \\
& +  \frac{2}{3} \frac{\left ( \textrm{Im} f_{em} (^3 P_0) - \textrm{Im} f_{em} (^3 P_2)  \right )}{m^4} \eta^\dagger \bfq' \cdot \bfSigma \xi \, \xi^\dagger \bfq \cdot \bfSigma \eta,
\end{align}
from which we can immediately read off the values of the matching coefficients
\begin{align}
\textrm{Im} f_{em} (^3 P_0) & = 3 \alpha^2 Q^4 \pi, \\
\textrm{Im} f_{em} (^3 P_2) & = \frac{4}{5}\alpha^2 Q^4 \pi,
\end{align}
that agree with the known results from the literature \cite{Bodwin1995,Petrelli1998,Ma2002,Mereghetti2006,Sang2015}.
\section{Summary}

We have presented new features and improvements in \textsc{FeynCalc} 9.0 and discussed cases in which \textsc{FeynCalc} can be used to obtain new results. Although the very first version of \textsc{FeynCalc} appeared almost 25 years ago, the development is still far from being complete. New developments in theoretical particle physics show possible directions in which \textsc{FeynCalc} can evolve. This includes better support for multi-loop calculations and determination of matching coefficients in effective field theories, but also built-in interfaces to other useful software tools and the ability to work with non-relativistic theories.

 Finally, we would like to observe that in the last two years some new general-purpose packages \cite{Wiebusch2014,Patel2015} for QFT  calculations were released, which follow the approach similar to that of \textsc{FeynCalc} and thus provide a comparable level of flexibility. This development shows that even in the age of fully automatic packages for 1-loop calculations, user-friendly, semi-automatic tools like \textsc{FeynCalc} are still in demand and employed in many interesting research projects.

\section*{Acknowledgments}

Two of the authors (RM and FO) would like to thank Daniel Wyler for his help and support in the earlier days of \textsc{FeynCalc}.

One of the authors (VS) wants to thank Hector Martinez Neira for bringing his attention to \textsc{FeynCalc} for the first time and his PhD supervisor Nora Brambilla for encouraging him to work in this direction. Simone Biondini is acknowledged for testing 1-loop tensor decompositions with \texttt{TID}. VS would also like to express his gratitude to Antonio Vairo, Christoph Bobeth, Thomas Hahn, Georg Weiglein, Sergey Larin, Claude Duhr, Matthias Steinhauser, Alexander Smirnov, Massimo Passera and Yu Jia for useful discussions and explanations. His work has been supported by the DFG and the NSFC through funds provided to the Sino-German CRC 110 ``Symmetries and the Emergence of Structure in QCD'', and by the DFG cluster of excellence ``Origin and structure of the universe'' (\url{www.universe-cluster.de}). 

Last but not least, all the authors would like to thank to the participants of the \textsc{FeynCalc} mailing list for their bug reports, feature requests, suggestions and encouragements.

\section*{References}






\bibliographystyle{bibstyle}
\bibliography{fcbiblio.bib}

\end{document}